\documentclass[journal]{IEEEtran}
\usepackage{color}
\usepackage{cite}

\ifCLASSINFOpdf
  \usepackage[pdftex]{graphicx}
\else
  \usepackage[dvips]{graphicx}
\fi
\usepackage{url}
\usepackage{amsmath,graphicx}
\usepackage{multirow,cite}
\usepackage{amsmath}
\usepackage[psamsfonts]{amssymb}
\usepackage{amsxtra}
\usepackage{threeparttable}
\usepackage{verbatim}
\usepackage[rawfloats=true]{floatrow} 
\restylefloat{figure}  
\usepackage{hhline}% http://ctan.org/pkg/hhline
\usepackage[rightcaption]{sidecap}
\usepackage {tikz}
\usetikzlibrary {positioning}
\usepackage{algorithmic,algorithm2e,float}
\usepackage[utf8]{inputenc}
\usepackage{fourier} 
\usepackage{array}
\usepackage{makecell}
\usepackage{diagbox}
\usepackage{cite}
\usepackage{caption}
\usepackage{subcaption}
\begin{document}
\title{An Overview of Voice Conversion and \\ its Challenges: From Statistical Modeling  to \\ Deep Learning}

\author{Berrak~Sisman,~\IEEEmembership{Member,~IEEE,}
        Junichi~Yamagishi,~\IEEEmembership{Senior Member,~IEEE,}
         Simon~King,~\IEEEmembership{Fellow,~IEEE,}
        and~Haizhou~Li,~\IEEEmembership{Fellow,~IEEE}% <-this % stops a space
\thanks{Berrak Sisman is with the Information Systems Technology and Design (ISTD) Pillar of Singapore University of Technology and Design (SUTD), Singapore. }% <-this % stops a space
\thanks{Junichi Yamagishi is with National Institute of Informatics, Japan and University of Edinburgh, United Kingdom. }% <-this % stops a space
\thanks{Simon King is with the University of Edinburgh, United Kingdom.}
\thanks{Haizhou Li is with the Department of Electrical and Computer Engineering, National University of Singapore.}
}

\maketitle

\begin{abstract}
Speaker identity is one of the important characteristics of human speech. In voice conversion, we change the speaker identity from one to another, while keeping the linguistic content unchanged. Voice conversion involves multiple speech processing techniques, such as speech analysis, spectral conversion, prosody conversion, speaker characterization, and vocoding. With the recent advances in theory and practice, we are now able to produce human-like voice quality with high speaker similarity. In this paper, we provide a comprehensive overview of the state-of-the-art of voice conversion techniques and their performance evaluation methods from the statistical approaches to deep learning, and discuss their promise and limitations. We will also report the recent Voice Conversion Challenges (VCC), the performance of the current state of technology, and provide a summary of the available resources for voice conversion research.
\end{abstract}

\begin{IEEEkeywords}
Voice conversion, speech analysis, speaker characterization, vocoding, voice conversion evaluation, voice conversion challenges.
\end{IEEEkeywords}

% \fi
\IEEEpeerreviewmaketitle

\section{Introduction}

Voice conversion (VC) is a significant aspect of artificial intelligence. It is the study of how to convert one's voice to sound like that of another without changing the linguistic content. Voice conversion belongs to a general technical field of speech synthesis, which converts text to speech or changes the properties of speech, for example, voice identity, emotion, and accents.  Stewart, a pioneer in speech synthesis, commented in 1922 \cite{Stewart1922}, the really difficult problem involved in the the artificial production of speech-sounds is not the making of a device which shall produce speech, but in the manipulation of the apparatus. As voice conversion is focused on the manipulation of voice identity in speech, it represents one of the challenging research problems in speech processing. 

There has been a continuous effort in quest for effective manipulation of speech properties since the debut of computer-based speech synthesis in the 1950s. The rapid development of digital signal processing in the 1970s greatly facilitated the control of the parameters for speech manipulation. While the original motivation of voice conversion could be simply novelty and curiosity, the technological advancements from statistical modeling to deep learning have made a major impact on many real-life applications, and benefited the consumers, such as personalized speech synthesis~\cite{kain1998spectral, zhang2019joint}, communication aids for the speech-impaired~\cite{voicebanking}, speaker de-identification~\cite{de-identification}, voice mimicry~\cite{wu_li_2014} and disguise~\cite{Huang2020DefendingYV}, and voice dubbing for movies.

In general, a speaker can be characterized by three factors that are 1) linguistic factors that are reflected in sentence structure, lexical choice, and idiolect; 2) supra-segmental factors such as the prosodic characteristics of a speech signal, and 3) segmental factors that are related to short term features, such as spectrum and formants. When the linguistic content is fixed, the supra-segment and the segmental factors are the relevant factors concerning speaker individuality. An effective voice conversion technique is expected to convert both the supra-segment and the segmental factors.  Despite much progress, voice conversion is still far from perfect. In this paper, we celebrate the technological advances, at the same time we expose their limitations. We will discuss the state-of-the-art technology from historical and technological perspectives. 

\begin{figure*}
  \centering
  \includegraphics[scale=0.7]{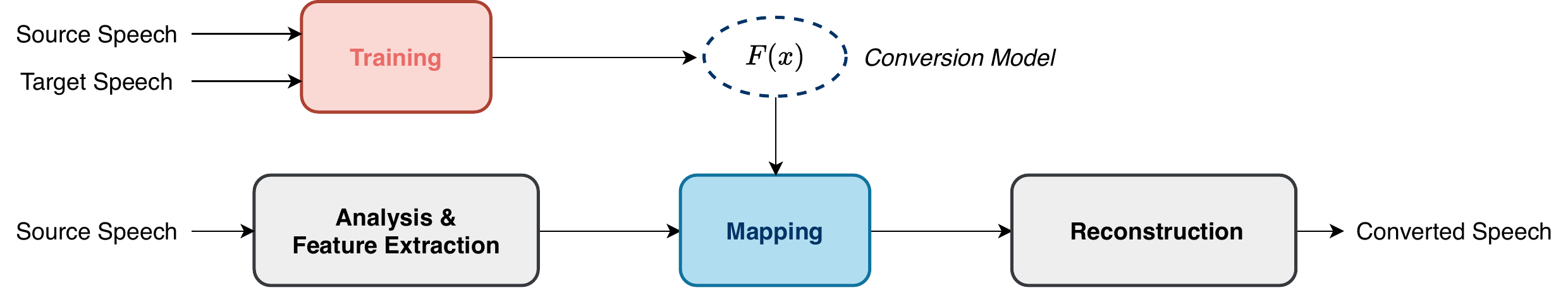}
 \caption{The typical flow of a voice conversion system. The pink box represents the training of the mapping function, while the blue box applies the mapping function at run-time, in a 3-step pipeline process $\mathcal{Y}=({\mathsf{R}}\circ {\mathsf{F}} \circ {\mathsf{A}}) (\mathcal{X})$.}
 \label{flow.pdf}
\end{figure*}

%%% parallel training data with statistical modeling 
A typical voice conversion pipeline  includes a speech analysis, mapping, and reconstruction modules as illustrated in Figure \ref{flow.pdf}, that is referred to as analysis-mapping-reconstruction pipeline. The speech analyzer decomposes the speech signals of a source speaker into features that represent supra-segmental and segmental information, and the mapping module changes them towards the target speaker, finally the reconstruction module re-synthesizes time-domain speech signals. The mapping module has taken the centre stage in many of the studies. These techniques can be categorized in different ways, for example, based on the use of training data -  parallel vs non-parallel, the type of statistical modeling technique - parametric vs non-parametric, the scope of optimization - frame level vs utterance level, and the workflow of conversion - direct mapping vs inter-lingual. Let's first give an account from the perspective of the use of training data. 

The early studies of voice conversion were focused on spectrum mapping using parallel training data, where  speech of the same linguistic content is available from both the source and target speaker, for example, vector quantization (VQ) \cite{Abe1988} and fuzzy vector quantization  \cite{Shikano1991}. With parallel data, one can align the two utterances using Dynamic Time Warping \cite{helander2008impact}. The statistical parametric approaches can benefit from more training data for improved performance, just to name a few, Gaussian mixture model \cite{Toda2007, Zen2008, Kobayashi2016},  partial least square regression \cite{Helander2010} and dynamic kernel partial least squares regression (DKPLS) \cite{Helander2012}.  

One of the successful statistical non-parametric techniques is based on non-negative matrix factorization (NMF) \cite{ nmf22} and it is known as the exemplar-based sparse representation technique  \cite{japaneseNMF, Wu2014, Aihara2016, Jin2016}. It requires a smaller amount of training data than the parametric techniques, and addresses well the over-smoothing problem. We note that the muffled sound effect occurs when the spectra are smoothed. The family of sparse representation techniques include phonetic sparse representation, group sparsity implementation\cite{Aihara2014, ccicsman2017sparse}, that greatly improved the voice quality on small parallel training dataset.

The studies on voice conversion towards non-parallel training data \cite{toda_cross, sundermann2006text, corsslingual-un2, corsslingual-fw, corsslingual-al2, corsslingual-al1} open up the opportunities for new applications. The challenge is how to establish the mapping  between non-parallel  source  and  target utterances. The INCA alignment technique by Erro et al. \cite{corsslingual-al2} represent one of the solutions to the non-parallel data alignment problem \cite{tao2010supervisory}. With the alignment techniques, one is able to extend the voice conversion techniques from parallel data to non-parallel data, such as the extension to DKPLS \cite{Silen2012} and speaker model alignment method \cite{Song2014}. 
%INCA alignment is based on an iteratively up-dated low-order auxiliary model that converts the set of source feature vectors to better match with the set of target feature vectors or vice versa.  The goal of INCA is to find the best matching frame pairs from the two non-parallel sets of speech spectral feature vector. 
%The idea can be extended to cross-lingual applications. 
%The evaluation of alternative alignment methods is discussed in \cite{tao2010supervisory}. 
%Moreover, an extended version of DKPLS-based conversion approach for non-parallel data with INCA alignment algorithm \cite{Silen2012} and speaker model alignment method \cite{Song2014} have been proposed and shown to achieve high-quality voice. 
Phonetic Posteriograms, or PPG-based \cite{lifasun}, approach represents another direction of research towards non-parallel training data. While the alignment technique doesn't use external resources, the PPG-based approach makes use of automatic speech recognizer to generate intermediate phonetic representation \cite{Hazen2009, Kintzley2011} as the inter-lingual between the speakers. Successful applications include Phonetic Sparse Representation\cite{ccicsman2017sparse}.

%We note that the traditional sparse representation approaches \cite{japaneseNMF, Wu2014} are not convenient for non-parallel data VC mainly because the activation matrix serves as a bridge between source and target speakers  across  the  parallel  data. When  parallel  training data are not available, we need to find ways to establish the mapping  between  source  and  target.   A  recent  study \cite{ccicsman2017sparse} solves this problem by using phonetic posteriorgrams (PPGs)  in  Phonetic  Sparse  Representation  (PSR)  to perform  such  a  mapping. PPGs represents  the  posterior  probability  of  each  phonetic  class for each speech frame \cite{Hazen2009, Kintzley2011}.   By converting all speaker’s  voice  into  PPGs,  one can estimate a  PPGs-derived  activation  matrix  to  serve  as  a  bridge  between  source  and target  speaker  at  run-time.  The proposed idea \cite{ccicsman2017sparse} is not only eliminates the need of parallel data, it also eminiates the need of source speaker's data during training. 
%%% deep learning with parallel training data 
Wu and Li \cite{wu_li_2014}, and Mohammadi and Kain \cite{mohammadi2017overview} provided an overview of voice conversion systems from the perspective of time alignment of speech features followed by feature mapping, that represents the statistical modeling school of thought. The advent of deep learning techniques represents an important technology milestone in the voice conversion research \cite{narendranath1995transformation}. It has not only greatly advanced the state-of-the-art, but also transformed the way we formulate the voice conversion research problems. It also opens up a new direction of research beyond the parallel and non-parallel data paradigm. Nonetheless, the studies on statistical modeling approaches have provided profound insights into many aspects of the research problems that serve as the foundation work of today's deep learning methodology. In this paper, we will give an overview of voice conversion research by providing a perspective that reveals the underlying design principles from statistical modeling to deep learning. 

Deep learning's contributions to voice conversion can be summarized in three areas. Firstly, it allows the mapping module to learn from a large amount of speech data, therefore, tremendously improves voice quality and similarity to target speaker. With neural networks, we see the mapping module as a nonlinear transformation function \cite{hornik1989multilayer}, that is trained from data \cite{laskar2012comparing, Nguyen2016}.  LSTM represents a successful implementation with parallel training data \cite{Sun2015}. Deep learning made a great impact on non-parallel data techniques. The joint use of DBLSTM and i-vector \cite{Wu}, KL divergence and DNN-based approach \cite{Xie2016}, variational auto-encoder \cite{Hsu2016}, average modeling \cite{Tianavmodel}, DBLSTM based Recurrent Neural Networks \cite{Sun2016, lifasun} and end-to-end Blow model \cite{serra2019blow} bring the voice quality to a new height.  More recently, Generative Adversarial Networks such as VAW-GAN \cite{hsu3}, CycleGAN \cite{kaneko2017parallel, cycleGAN2, steal-vc}, and many-to-many mapping with StarGAN \cite{stargan-vc} further advance the state-of-the-art.  %We note that Generative Adversarial Networks have also been shown to be very effective in translating an image from a source domain to a target domain in the absence of parallel data \cite{image1, image2, image3}, that motivates the studies in voice conversion.

Secondly, deep learning has created a profound impact on vocoding technology. Speech analysis and reconstruction modules are typically implemented using a traditional parametric vocoder \cite{airaksinen2018comparison, Toda2007, Zen2008, Kobayashi2016}. The parameters of such vocoders are manually tuned according to some over-simplified assumptions in signal processing. As a result, the parametric vocoders offer a suboptimal solution.
%and the difficulty to accurately estimate the features from the speech signal.
%For example, Griffin-Lim is not capable of generating speech with acceptable naturalness due to the lack of phase information in the short-time Fourier transform (STFT).
Neural vocoder is a neural network that learns to reconstruct an audio waveform from acoustic features \cite{WangLTJY18_NeuralVocoders_ICASSP}. For the first time, neural vocoder becomes trainable and data-driven. WaveNet vocoder \cite{Hayashi2017AnIO} represents one of the popular neural vocoders, that directly estimates waveform samples from the input feature vectors. It has been studied intensively, for example, speaker dependent and independent WaveNet vocoder \cite{vc_wavenet, Hayashi2017AnIO}, quasi-periodic WaveNet vocoder \cite{wu2019quasi, wu2019statistical}, adaptive WaveNet vocoder with GANs \cite{sisman2018adaptive}, factorized WaveNet vocoder~\cite{hongqiang2019asru}, and refined WaveNet vocoder with VAEs \cite{huang2019refined} that are known for their natural sounding voice quality. WaveNet vocoder has been widely adopted in traditional voice conversion pipelines, such as GMM \cite{vc_wavenet}, sparse representation \cite{berrak_is18, berrak-journal} systems. Other successful neural vocoders include WaveRNN vocoder\cite{kalchbrenner2018efficient}, WaveGlow \cite{PrengerVC19_WaveGlow_ICASSP} and FloWaveNet \cite{kim2019flowavenet} that are excellent vocoders in their own right.  
%The WaveNet approach transforms the vocoder design into a data-driven learnable process. Through the learning, the network is expected to capture the dynamics of the complex mechanics of the human speech generation process. There have been reported work on voice conversion with WaveNet vocoder such as GMM \cite{vc_wavenet}, sparse representation \cite{berrak_is18, berrak-journal}, quasi-periodic WaveNet vocoder \cite{wu2019quasi, wu2019statistical}, adaptive WaveNet vocoder with GANs \cite{sisman2018adaptive} and refined WaveNet vocoder with VAEs \cite{huang2019refined}. 

%% waveRNN and waveGlow
%More recently, speaker independent WaveRNN-based neural vocoder \cite{kalchbrenner2018efficient} became popular as it can generate human-like voices regardless of whether the input spectrogram comes from  a  speaker  or  style  seen  during  training  or  from  an out-of-domain  scenario  when  the  recording  conditions  are studio-quality \cite{lorenzo2019towards, govalkar2019comparison}. Another well-known neural vocoder that is known to achieve high-quality performance is WaveGlow \cite{PrengerVC19_WaveGlow_ICASSP}. WaveGlow is a flow-based network capable of generating high quality speech from mel-spectrograms. WaveGlow combines insights from Glow and WaveNet in order to provide fast, efficient and high-quality audio synthesis, without the need for auto-regression. WaveGlow is implemented using only a single network, trained using only a single cost function: maximizing the likelihood of the training data, which makes the training procedure simple and stable.

Thirdly, deep learning represents a departure from the traditional analysis-mapping-reconstruction pipeline.
All the above techniques largely follow the  voice   conversion  pipeline as in Figure 1. As neural vocoder is trainable, it can be trained jointly with mapping module \cite{sisman2018adaptive} and even with analysis module to become end-to-end solution \cite{van2016wavenet}.

%% voice conversion challenges
Voice conversion research used to be a niche area in speech synthesis. However, it has become a major topic in recent years. In the 45th International Conference on Acoustics, Speech, and Signal Processing (ICASSP 2020), voice conversion papers represent more than one-third of the papers under the speech synthesis category.  The growth of research community was accelerated by collaborative activities across academia and industry, such as voice conversion challenge (VCC) 2016, which was first launched \cite{toda2016voice, wester2016multidimensional, wester2016analysis} at INTERSPEECH 2016.  VCC 2016 is focused on the most basic voice conversion task, that is voice conversion for parallel training data recorded in acoustic studio. %The objective of the challenge was to use the same data to understand the unsolved  problems  and  challenges  faced  by  current voice conversion techniques. 
It establishes the evaluation methodology and protocol for performance benchmarking, that are adopted widely in the community.
%In 2018, the second edition of VCC was launched \cite{lorenzo2018voice, lorenzo2018voice2, tobing2018nu}. 
VCC 2018 \cite{lorenzo2018voice, lorenzo2018voice2, tobing2018nu} proposes a non-parallel training data challenge, and also connects voice conversion with anti-spoofing of speaker verification studies. VCC 2020 puts forward a cross-lingual voice conversion challenge for the first time. We will provide an overview of the series of challenges and the publicly available resources in this paper.

%differs from VCC 2016 in three ways: 1) the amount of speech data used for training is reduced to half; 2) a more challenging task where participants are allowed to use a non-parallel database in which source and target speakers read out different sets of utterances; 3) Evaluation of both parallel and non-parallel voice conversion systems via the same large-scale crowd sourcing listening test. Voice conversion  is  also  known  as  an  advanced  presentation attack method against automatic speaker verification systems. This challenge also acts as a bridge between the automatic speaker verification (ASV) and voice conversion communities. we will provide an overview of VCC 2016, 2018, and 2020 challenges.

%% overall plan 
This paper is organized as follows: In Section II, we present the typical flow of voice conversion that includes feature extraction, feature mapping and waveform generation. In Section III, we study the statistical modeling for voice conversion with parallel training data. In Section IV, we study statistical modeling for voice conversion without parallel training data. In Section V, we study the deep learning approaches for voice conversion with parallel training data, and beyond parallel training data. In Section VI, we explain the evaluation techniques for voice conversion. In Section VII and VIII,  we summarize the series of voice conversion challenges, and publicly available research resources for voice conversion. We conclude in Section IX.

\section{Typical flow of voice conversion}

The goal of voice conversion is to modify a source speaker’s voice to sound as if it is produced by a target speaker. In other words, a voice conversion system only modifies the speaker-dependent characteristics of speech, such as formants, fundamental frequency (F0), intonation, intensity and duration,  while carrying over the speaker-independent speech content.

The core module of a voice conversion system performs the conversion function. Let's denote the source and target speech signals as $\mathcal{X}$ and $\mathcal{Y}$ respectively. As will be discussed later, voice conversion is typically applied to some intermediate representation of speech, or speech feature, that characterizes a speech frame. Let's denote the source and target speech features as $\textbf{x}$ and $\textbf{y}$. The conversion function can be formulated as follows, 

\begin{align}
    {\textbf{y}}=\mathsf{F}(\textbf{x})
\end{align}
where $\mathsf{F}(\cdot)$ is also called frame-wise mapping function in rest of this paper. 
% Voice conversion approaches modify a source speaker's voice to sound like it was produced by the target speaker by converting spectral and prosody features. We note that the voice conversion technology can be implemented for both mono-lingual and cross-lingual data of source and target speakers. Regardless of mono-lingual or cross-lingual training data, 
As illustrated in Figure \ref{flow.pdf}, a typical voice conversion framework is implemented in three steps: 1) speech analysis, 2) feature mapping, and 3) speech reconstruction, that we call the analysis-mapping-reconstruction pipeline. We discuss in detail next.  

\subsection{Speech Analysis and Reconstruction}
The speech analysis and reconstruction are two crucial processes in the 3-step pipeline. The goal of speech analysis is to decompose speech signals into some form of intermediate representation   for effective manipulation or modification with respect to the acoustic properties of speech. There have been many useful intermediate representation techniques that were initially studied for speech communication, and speech synthesis. They become handy for voice conversion. In general, the techniques can be categorized into model-based representations, and signal-based representations.  

In model-based representation, we assume that speech signal is generated according to a underlying physical model, such as source-filter model, and 
express a frame of speech signal as a set of model parameters. By modifying the parameters, we manipulate the input speech. In signal-based representation, we don't assume any models, but rather represent speech as a composition of controllable elements in time domain or frequency domain. Let's denote the intermediate representation for source speaker as \textbf{x}, speech analysis can be described by a function,

\begin{equation}
    \textbf{x}=\mathsf{A}(\mathcal{X})
\end{equation}

Speech reconstruction  can be seen as an inverse function of the speech analysis, that operates on the modified parameters and generates an audible speech signal. It works with speech analysis in tandem. For example, A  vocoder \cite{airaksinen2018comparison} is used to express a speech frame with a set of controllable parameters that can be converted back into a speech waveform. A Griffin-Lim algorithm is used to reconstruct a speech signal from a modified short-time Fourier transform after amplitude modification \cite{griffin1984signal}. As the output speech quality is affected by the speech reconstruction process,  speech reconstruction is also one of the important topics in voice conversion research.  
Let's denote the modified intermediate representation and the reconstructed speech signal for target speaker as ${\textbf{y}}$ and $\mathcal{Y}=\mathsf{R}(\textbf{y})$, voice conversion can be described by a composition of three functions,
\begin{equation}
\begin{split}
 %  \mathcal{Y}&={\mathsf{R}}({\textbf{y}})\\
   \mathcal{Y}&=({\mathsf{R}}\circ {\mathsf{F}} \circ {\mathsf{A}}) (\mathcal{X})\\
   &=\mathsf{C}(\mathcal{X})
\end{split}
\end{equation}  
\noindent{that represents the typical flow of a voice conversion system as a 3-step pipeline. As the mapping is applied  frame-by-frame, the number of converted speech features \textbf{y} is the same as that of the source speech features \textbf{x} if speech duration is not modified in the process.}
%The design of the modules is motivated by speech production process, for example, source-filter model is often adopted in practice.   When choosing the speech production model, the speech analysis module is expected to be able to separate a speech signal into several mutually independent representation components, such as spectrums, fundamental frequencies (F0) and excitation, for flexible modifications without introducing artifacts. This requirement enables the representation components to be modified independently. 

While speech analysis and reconstruction make possible voice conversion, just like other signal processing techniques, they inevitably also introduce artifacts. Many studies were devoted to minimize such artifacts. We next discuss the most commonly used speech analysis and reconstruction techniques in voice conversion. 

\subsubsection{Signal-based Representation} 

Pitch Synchronous OverLap and Add (PSOLA)
is an example of signal-based representation techniques. It decomposes a speech signal into overlapping speech segments \cite{moulines1990pitch}, each of which represents one of the successive pitch periods of the speech signal. By overlap-and-adding these speech segments with a different pitch periods, we can reconstruct the speech signal of a different intonation.
As PSOLA operates directly on the time-domain speech signal \cite{moulines1990pitch}, the analysis and reconstruction do not introduce significant artifacts. While PSOLA technique is effective for modification of fundamental frequency of speech signals, it suffers from several inherent limitations \cite{valbret1992voice, arslan1999speaker}. For example, unvoiced speech signal is not periodic, and the manipulation of time-domain signal not straightforward.

Harmonic plus Noise Model (HNM) represents another signal-based representation approach. It works under the assumption that a speech signal can be represented as a harmonic component plus a noise component that is delimited by the so-called maximum voiced frequency \cite{stylianou2001applying}. The harmonic component is modeled as the sum of harmonic sinusoids up to the maximum voiced frequency, while the noise component is modeled as Gaussian noise filtered by a time-varying autoregressive filter. As HNM decomposition is represented by some controllable parameters, it allows for easy modification speech \cite{stylianou1998system, erro2007weighted}.  

\subsubsection{Model-based Representation} 
The model-based technique assumes that the input signal can be mathematically represented by a model whose parameters vary with time. A typical example is the source-filter model that represents a speech signal as the outcome of an excitation of the larynx (source) modulated by a transfer (filter) function determined by the shape of the supralaryngeal vocal tract. A  vocoder, a short form of voice coder, was initially  developed to minimize the amount of data that are transmitted for voice communication. It encodes speech into slowly changing control parameters, such as linear predictive coding and mel-log spectrum approximation \cite{imai1983mel}, that describe the filter, and re-synthesizes the speech signal with the source information at the receiving end. In voice conversion, we convert the speech signals from a source speaker to mimic the target speaker by modifying the controllable parameters.

The majority of vocoders are designed based on some form of the source-filter model of speech production, such as mixed excitation with a spectral envelope, and glottal vocoders \cite{8357921}. STRAIGHT or ``Speech Transformation and Representation using Adaptive Interpolation of weiGHTed spectrum" is one of the popular vocoders in speech synthesis and voice conversion \cite{kawahara1999restructuring}. It decomposes a speech signal into: 1) a smooth spectrogram which is free from periodicity in time and frequency; 2) a fundamental frequency (F0) contour which is estimated using a fixed-point algorithm; and 3) a time-frequency periodicity map which captures the spectral shape of the noise and its temporal envelope. STRAIGHT is widely used in voice conversion because its parametric representation facilitates the statistical modeling of speech, that allows for easy manipulation of speech \cite{Toda2007, desai2009voice, berrak_is18_2}. %, that is typical example of parametric vocoders.  

Parametric vocoders are widely adopted for analysis and reconstruction of speech in voice conversion studies \cite{Abe1988, Shikano1991, Toda2007, Zen2008, kaneko2017parallel, hsu1, hsu2, hsu3}, and continue to play a major role today\cite{japaneseNMF, Aihara2014, ccicsman2017sparse}. The traditional parametric vocoders are designed to approximate the complex mechanics of the human speech production under certain simplified assumptions. For example, the interaction between F0 and formant structure is ignored, and the original phase structure is discarded \cite{furui}. The assumption of stationary process in the short-time window, and time-invariant linear filter, also give rise to “robotic” and “buzzy” voice. 
Such problems become more serious in voice conversion as we modify both F0 and the formant structure of speech among others at the same time. We believe that vocoding can be improved by considering the interaction between the parameters.

\subsubsection{WaveNet Vocoder}
Deep learning offers a solution to some of the inherent problems of parametric vocoders.  WaveNet \cite{van2016wavenet} is a deep neural network that learns to generate high quality time-domain waveform. As it doesn't assume any mathematical model, it is a data-driven solution that requires a large amount of training data. 

The joint probability of a waveform $\mathcal{X}= {x_1, x_2, ..., x_N}$ can be factorized as a product of conditional probabilities. 
\begin{align}
p(\mathcal{X})=\prod_{n=1}^{N}p(x_n|x_1, x_2, ..., x_{n-1})
\label{eqn:wavenet}
\end{align}
A WaveNet is constructed with many residual blocks, each of which consists of  $2 \times 1$ dilated causal convolutions, a gated activation function and $1 \times 1$  convolutions. With additional auxiliary features $h$, WaveNet can also model conditional distribution $p(x|h)$\cite{van2016wavenet}. Eq. (\ref{eqn:wavenet}) can then be written as follows:
\begin{align}
p(\mathcal{X}|h)=\prod_{n=1}^{N}p(x_n|x_1, x_2, ..., x_{n-1},h)
\end{align}
A typical parametric vocoder performs both analysis and reconstruction of speech. However, most of today's WaveNet vocoders only cover the function of speech reconstruction. It takes some intermediate representations of speech as the input auxiliary features, and generate speech waveform as the output.
%There have been promising studies that use parametric vocoding parameters \cite{vc_wavenet, huang2019refined, wu2019statistical, Hayashi2017AnIO, tacotron2, berrak_is18, chen2018high, adiga2018use, zhao2018wasserstein}, and phonetic posteriogram (PPG) \cite{tian2019speaker, lu2019compact, du2019wavenet, liu2019jointly} as the intermediate representations in WaveNet vocoding. We note that in the original WaveNet \cite{van2016wavenet}, linguistic features and/or speaker codes are conditioned to generate speech samples  according to the given text input while keeping specific speaker characteristics.
WaveNet vocoder \cite{Hayashi2017AnIO} outperforms remarkably the traditional parametric vocoders in terms of sound quality. Not only can it learn the relationship between input features and output waveform, but also it learns the interaction among the input features. It has been successfully adopted as part of the state-of-the-art speech synthesis \cite{tacotron2, zhang2019joint, liu2019teacher, hanzlivcek2018wavenet, arik2017deep} and voice conversion \cite{sisman2019machine, berrak-journal, vc_wavenet, huang2019refined, wu2019statistical, Hayashi2017AnIO, tacotron2, berrak_is18, chen2018high, adiga2018use, zhao2018wasserstein, tian2019speaker, lu2019compact, du2019wavenet, liu2019jointly} systems.  

There have been promising studies on using vocoding parameters as the intermediate representations in WaveNet vocoding. A speaker independent WaveNet vocoder~\cite{Hayashi2017AnIO} is studied by utilizing the STRAIGHT vocoding parameters, such as F0, aperiodicity, and spectrum as the inputs of WaveNet. In this way, WaveNet learns a sample-by-sample correspondence between the time-domain waveform and the input vocoding parameters. When such a WaveNet vocoder is trained on speech signals from a large speaker population, we obtain a speaker independent vocoder\cite{Hayashi2017AnIO}. By adapting the speaker independent WaveNet vocoder with speaker specific data, we obtain a speaker dependent vocoder that generates personalized voice output \cite{sisman2018adaptive, huang2019refined}. The study on WaveNet vocoder also opens up opportunities for the use of other non-vocoding parameters as the input. For example, a recent study adopts phonetic posteriogram (PPG) in WaveNet vocoding with promising results in voice conversion with non-parallel training data  \cite{tian2019speaker, lu2019compact, du2019wavenet, liu2019jointly}. Another study adopts latent code of autoencoder and speaker embedding as the speech representation for WaveNet vocoder~\cite{wavenet-vae2019}.

\subsubsection{Recent Progress on Neural  Vocoders}
More recently, speaker independent WaveRNN-based neural vocoder \cite{kalchbrenner2018efficient} became popular as it can generate human-like voices from both in-domain and out-of-domain spectrogram~\cite{lorenzo2019towards, govalkar2019comparison, yi2019singing}. Another well-known neural vocoder that achieves high-quality synthesis performance is WaveGlow \cite{PrengerVC19_WaveGlow_ICASSP}. WaveGlow is a flow-based network capable of generating high quality speech from mel-spectrogram \cite{okamoto2019real}. WaveGlow benefits from the best of Glow and WaveNet so as to provide fast, efficient and high-quality audio synthesis, without the need for auto-regression. We note that WaveGlow is implemented using only a single network with a single cost function, that is to maximize the likelihood of the training data, which makes the training procedure simple and stable \cite{maiti2019parametric}. 

%Neural waveform models such as WaveNet have demonstrated better performance than conventional vocoders for statistical parametric speech synthesis \cite{van2016wavenet,Hayashi2017AnIO }. 
WaveNet~\cite{van2016wavenet} uses an auto-regressive (AR) approach to model the distribution of waveform sampling points, that incurs a high computational cost. As an alternative to auto-regression, a neural source-filter (NSF) waveform modeling framework is proposed~\cite{wang2019neural, wang2019neuralarxiv}. We note that NSF is straightforward to train and fast to generate waveform. It is reported 100 times faster than WaveNet vocoder, and yet achieving comparable voice quality on a large speech corpus \cite{wang2019neural2}. 

More recently, Parallel WaveGAN \cite{yamamoto2020parallel} has also been proposed to generate high-quality voice using a generative adversarial network. Parallel WaveGAN is a distillation-free and fast waveform generation method, where a non-autoregressive WaveNet is trained by jointly optimizing multi-resolution spectrogram and adversarial loss functions. We note that Parallel WaveGAN is able to generate high-fidelity speech even with its compact architecture. We note that generating coherent raw audio waveforms with GANs is challenging. Another GAN method for generating high quality audio waveform is known as MelGAN \cite{kumar2019melgan}. MelGAN shows the effectiveness of GAN-based approaches for high quality mel-spectrogram inversion in speech synthesis, music domain translation and unconditional music synthesis.

\subsection{Feature Extraction}
With speech analysis, we derive vocoding parameters that usually contains spectral and prosodic components to represent the input speech. The vocoding parameters characterize the speech in a way that we can reconstruct the speech signal later on after transmission. This is particularly important in speech communication.  However, such vocoding parameters may not be the best for transformation of voice identity. More often, the vocoding parameters are further transformed into speech features, that we call feature extraction in Figure 1, for more  effective modification  of the acoustic properties in voice conversion. 

For the spectral component, feature extraction aims to derive low-dimensional representations from the high-dimensional raw spectra. Generally speaking, the spectral features are 
 be able to represent the speaker individuality well. The feature not only fit the spectral envelope well, but also be able to be converted back to spectral envelope. They should have good interpolation properties that allow for flexible modification. 

The magnitude spectrum can be warped to Mel or Bark frequency scale that are perceptually meaningful for voice conversion. It can also be transformed into cepstral domain using a finite number
of coefficients using the Discrete Cosine Transform of log-magnitude. Cepstral coefficients are less correlated. In this way, high dimension magnitude spectrum is transformed to lower dimension feature representation. The commonly used speech features include Mel-cepstral coefficients (MCEP),  linear predictive cepstral coefficients (LPCC), and line   spectral frequencies  (LSF). Typically, a speech frame is represented by a feature vector.

Short-time analysis has been the most practical way of speech analysis. Unfortunately it inherently ignores the temporal context of speech, that is crucial in voice conversion. Many studies have shown that multiple frames\cite{Wu2014,Tian2016}, dynamic features\cite{berrak-journal}, and phonetic segments serve as effective features in feature mapping. 

For the prosodic component, feature extraction can be used to decompose prosodic signal, such as fundamental frequency (F0), aperiodicity (AP), and energy contours, into speaker dependent and independent parameters\cite{berrak_is18_2}. In this way, we can carry over the speaker independent prosodic patterns, while converting speaker dependent ones during the feature mapping. 

\subsection{Feature Mapping}
In the typical flow of voice conversion, feature mapping performs the modification of speech features from source to target speaker. Spectral mapping seeks to change the voice timbre, while prosody conversion seeks to modify the prosody features, such as fundamental frequency, intonation and duration. So far, spectral mapping remains the center of many voice conversion studies. 

During training, we learn the mapping function,  $\mathsf{F}(\cdot)$ in Eq.(1), from training data. At run time inference, the mapping function transforms the acoustic features. A large part of this paper is devoted to the study of the mapping function. In Section III, we will discuss the traditional statistical modeling techniques with parallel training data. In Section IV, we will review the statistical modeling techniques that do not require parallel training data. In Section V, we will introduce a number of deep learning approaches, which includes 1) parallel training data of paired speakers; and 2) beyond parallel data of paired speakers. %We will also study the transition from statistical models to deep learning.  

\section{Statistical Modeling for Voice Conversion with Parallel Training Data}

Most of the traditional voice conversion techniques assume availability of parallel training data. In other words, the mapping function is trained on paired utterances of the same linguistic content spoken by source and target speaker. Voice conversion studies started with statistical approaches\cite{kuwabara1995acoustic} in late 1980s, that can be grouped into parametric and non-parametric mapping techniques. Parametric techniques makes assumptions about the underlying statistical distributions of speech features and their mapping. Non-parametric ones make fewer assumptions about the data, but seek to fit the training data with the best mapping function, while maintaining some ability to generalize to unseen data.

Parametric techniques, such as Gaussian mixture model (GMM) \cite{stylianou1998continuous},  Dynamic Kernel Partial Least Square Regression,  PSOLA mapping technique \cite{valbret1992voice}, represent a great success in the recent past. The vector quantization approach to voice conversion is a typical non-parametric technique. It maps codewords between source and target codebooks\cite{Abe1988}. In this method, a source feature vector is approximated by the nearest codeword in the source codebook, and mappped to the corresponding codeword in the target codebook. To reduce the  quantization  error, fuzzy vector quantization was  studied \cite{Shikano1991, matsumoto1993unsupervised}, where continuous weights for individual clusters are determined at each frame according to the source feature vector. The converted feature vector is defined as a weighted sum of the centroid vectors of the mapping codebook. Recently, the non-negative factorization approach marks a successful non-parametric implementation. 

\begin{figure*}
  \centering
  \includegraphics[scale=0.55]{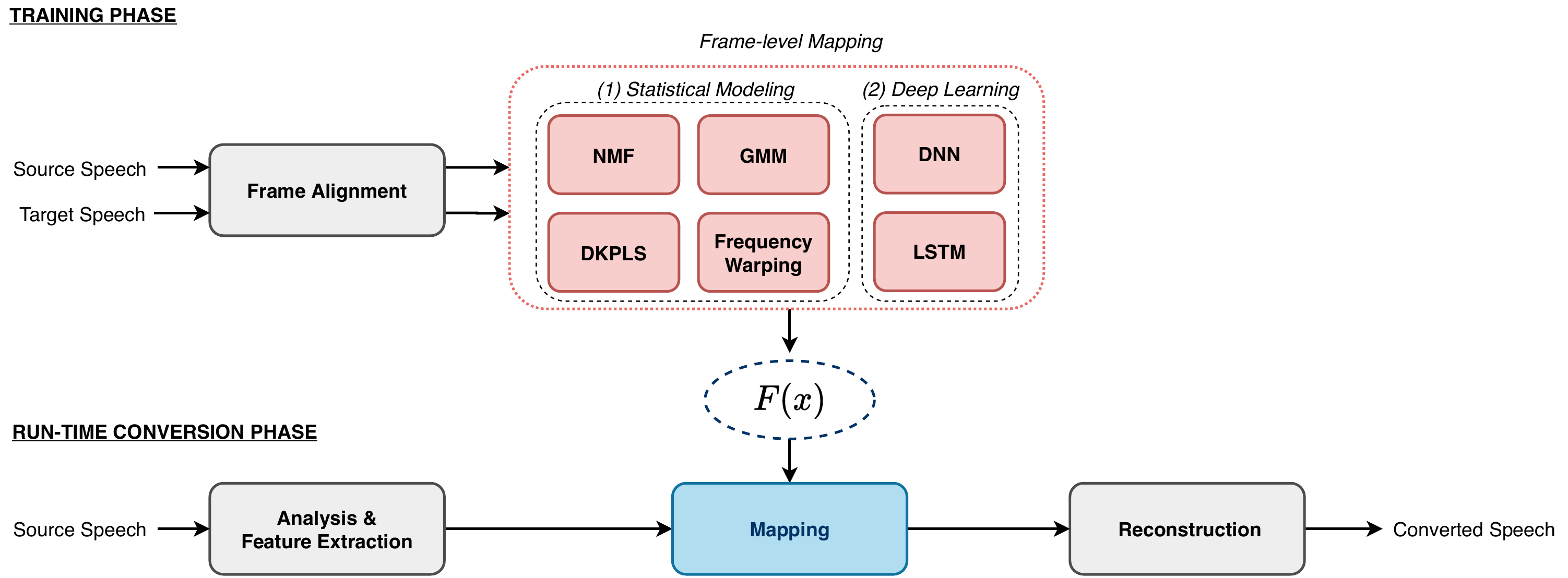}
 \caption{Training and run-time inference of voice conversion with parallel training data under the frame-level mapping paradigm. The pink boxes represent the training algorithms of the models that result in the mapping function $F(x)$ in blue box for run-time inference. Dotted box (1) includes examples of statistical approaches, and (2) includes examples of deep learning approaches. }
 \label{statistical-withparallel}
\end{figure*}

We will discuss a typical frame-level mapping paradigm under the assumption of parallel training data, as illustrated in Figure \ref{statistical-withparallel}. During the training phase, given parallel training data from a source speaker $\textbf{x}$ and a target speaker $\textbf{y}$, frame alignment is performed to align the source speech vectors and target speech vectors to obtain the paired speech feature vector $\textbf{z}=\{ \textbf{x}, \textbf{y} \}$. Dynamic time warping is feature-based alignment technique that is commonly used. Speech recognizer, that is equipped with phonetic knowledge, can also be used to perform model-based alignment. Frame alignment has been well studied in speech processing. In voice conversion, a large body of literature has been devoted to the design of frame-level mapping function. 

\subsection{Gaussian Mixture Models}
In Gaussian mixture modeling (GMM) approach to voice conversion \cite{stylianou1998continuous},
we represent  the relationship between  two  sets  of  spectral  envelopes, from source and target speakers, using a Gaussian mixture model. The Gaussian mixture model is a continuous parametric function, that is trained to model the spectral mapping.  In \cite{stylianou1998continuous}, harmonic plus noise (HNM) features are used in the feature mapping,  which  allows for high-quality  modifications of speech signals. The GMM approach is seen as an extension to the vector quantization approach\cite{Abe1988, Shikano1991}, that results in improved voice quality. However, the speech  quality  is  affected  by  some  factors,  e.g., spectral movement with inappropriate dynamic characteristics
caused by the frame-by-frame conversion process, and excessive smoothing of converted spectra \cite{toda2001voice, toda2000voice, toda2005spectral}. 

To address the frame-by-frame conversion issue, a maximum likelihood estimation technique was studied to model the spectral parameter trajectory  \cite{Toda2007}.  This technique aims to estimate an appropriate spectrum sequence using dynamic acoustic features. To address the over-smoothing issue, or the muffled effect, joint density Gaussian
mixture model (JD-GMM) was studied \cite{Toda2007, kain1998spectral} to jointly model the sequences of spectral features and their variances using maximum likelihood estimation, that increases the global variance  of the spectral features. 
The JD-GMM method involves two phases: off-line training and run-time conversion phases. 
%During the training phase, given parallel training data from a source speaker $\textbf{x}$ and a target speaker $\textbf{y}$, dynamic time warping (DTW) algorithm is used to align the source speech vectors and target speech vectors to obtain the paired speech feature vector $\textbf{z}=\{ \textbf{x}, \textbf{y} \}$. 
During the training phase, Gaussian mixture model (GMM) is adopted to model the joint probability density $p(\textbf{z})$ of the paired feature vector sequence $\textbf{z}=\{ \textbf{x}, \textbf{y} \}$, which represents the joint distribution of source speech $\textbf{x}$ and target speech $\textbf{y}$:
\begin{align}
    p(\textbf{z})=\sum_{k=1}^{K}w_{k}^{({z})}\mathcal{N}\left(\textbf{z} | \mu_{k}^{z}, \Sigma_{k}^{(z)} \right) 
\end{align}
\begin{align}
\mu_{k}^{z} = \begin{bmatrix} 
\mu_{k}^{x} \\ \\
\mu_{k}^{y}\\
\end{bmatrix}, \Sigma_{k}^{(z)}=\begin{bmatrix} 
\Sigma_{k}^{(xx)} & \Sigma_{k}^{(xy)} \\ \\
\Sigma_{k}^{(yx)} & \Sigma_{k}^{(yy)} \nonumber
\end{bmatrix}
\end{align}
where $K$ is the number of Gaussian components, $w_{k}$ is the weight of each Gaussian, $\mu_{k}^{z}$ and $\Sigma_{k}^{(z)}$ are the mean vector and the covariance matrix of the $k$th Gaussian component $ \mathcal{N}\left(\textbf{z} | \mu_{k}^{z}, \Sigma_{k}^{(z)} \right)$, respectively. To estimate the model parameters of the JD-GMM, expectation-maximization (EM) algorithm \cite{moon1996expectation, do2008expectation, xuan2001algorithms, gupta2011theory} is used to maximize likelihood on the training data. During the run-time conversion phase, JD-GMM model parameters are used to estimate the conversion function. We note that JD-GMM training method provides estimates of the model parameters robustly, especially when the amount of training data is limited.

A post-filter based on modulation spectrum modification is found useful to address the inherent over-smoothing issue in statistical modeling~\cite{takamichi2014modulation}, such as GMM approach, which effectively compensates the global variance. The GMM approach is a parametric solution \cite{ohtani2006maximum, kawanami2003gmm, aihara2012gmm, hwang2013incorporating, zorilua2012improving}. It represents a successful statistical modeling technique that works well with parallel training data.

\subsection{Dynamic Kernel Partial Least Squares}
%https://www.isca-speech.org/archive/interspeech_2014/i14_2318.html
The family of parametric techniques also include linear  \cite{valbret1992voice, arslan1999speaker} or non-linear  mapping functions. With the local mapping functions, each frame of speech is typically transformed independently from the neighboring frames, which causes temporal discontinuities to the output~\cite{arslan1999speaker}. 

To take into account the time-dependency between speech features, a dynamic kernel partial least squares (DKPLS) technique was studied \cite{Helander2012}. This  method  is  based  on  a  kernel transformation  of  the  source  features  to  allow  non-linear  modeling,  and  concatenation adjacent frames to model the dynamics. The non-linear  transformation takes  advantage  of the  global  properties  of  the  data that GMM approach doesn't. It was reported that DKPLS outperforms GMM approach \cite{stylianou1998continuous} in terms of voice quality. This method is simple and efficient, and does not require  massive tuning. More recently, DKPLS-based approaches are studied to overcome the over-fitting and over-smoothing problems by feature combination strategy~\cite{ghorbandoost2015voice}. 

While statistical modeling for the mapping of spectral features has been well studied, conversion of prosody is often achieved by simply shifting and scaling F0, which is not sufficient for high-quality voice conversion. Hierarchical modeling of prosody, for different linguistic units at several distinct temporal scales, represents an advanced technique for prosody conversion \cite{ribeiro2015perceptual, ribeiro2016wavelet, berrak_is18_2, wang2008multi}. DKPLS has created a platform for multi-scale prosody conversion through wavelet transform \cite{Sanchez2014} that shows significant improvement in naturalness over the F0 shifting and scaling technique.

%% Have a paragraph for DKPLS for how people use it (2-3 papers)...

\subsection{Frequency Warping}
Parametric techniques, such as GMM \cite{stylianou1998continuous} and DKPLS \cite{Helander2012}, usually suffer from over-smoothing because they use the minimum mean square error \cite{desai2009voice} or the maximum likelihood \cite{Toda2007} function as the optimization criterion. As a result, the system produces acoustic features that represent statistical average, and fails to capture the desired details of temporal and spectral dynamics. 

Additionally, parametric techniques generally employ low-dimensional features, as discussed in Section II.B, such as the Mel-cepstral coefficients (MCEP) or line spectral frequencies (LSF) to avoid the curse of dimensionality. The low dimensional features, however, are doomed to lose spectral details because they have low-resolution.  Statistical averaging and low-resolution features both lead to the muffled effect of output speech \cite{erro2009voice}. 

To preserve the necessary spectral details during conversion, a number of frequency warping-based methods were introduced. The frequency warping technique directly transforms the high resolution source spectrum to that of the target speaker through a frequency warping function. In recent literature, the warping function is either realized by a single parameter, such as VTLN-based approaches \cite{sundermann2003vtln, corsslingual-fw, eichner2004voice, pvribilova2006non, vich2012pitch}, or represented as a piecewise linear function \cite{erro2009voice,valbret1992voice, godoy2011voice}, which has become a mainstream solution.

The goal of piecewise linear warping function is to align a set of frequencies between the source and target spectrum by minimizing the spectral distance or maximizing the correlation between the converted and target spectrum. More recently, the parametric frequency warping technique was incorporated with a non-parametric exemplar-based technique, that achieves good performance \cite{Tian2016}.

\subsection{Non-negative Matrix Factorization}

Non-negative matrix factorization (NMF) \cite{Lee2001} is an effective data mining technique that has been widely used, especially for reconstruction of high quality signals, such as in speech enhancement \cite{nmf_enhancement,Mohammadiha2013}, speech de-noising \cite{Akarsh2015,Wilson}, noise and speech estimation \cite{nmf_Sun2015}. It factorizes a matrix into two matrices, a dictionary and an activation matrix, with the property that all three matrices have no negative elements.  The NMF-based techniques are shown effective in voice conversion with very limited training data. It marks a major progress of non-parametric approach to voice conversion since  vector quantization technique was introduced. Successful implementation includes non-negative spectrogram deconvolution \cite{Wu2013},  locally linear embedding (LLE) \cite{Wu2016}, and unit selection \cite{Jin2016}. 
In NMF-based approaches, a target spectrogram is constructed as a linear combination of exemplars. Therefore, over-smoothing problem can also arise. To overcome the over-smoothing problem, several effective techniques were developed, that we summarize next.
\begin{figure}[t]
  \centering
  \includegraphics[scale=0.7]{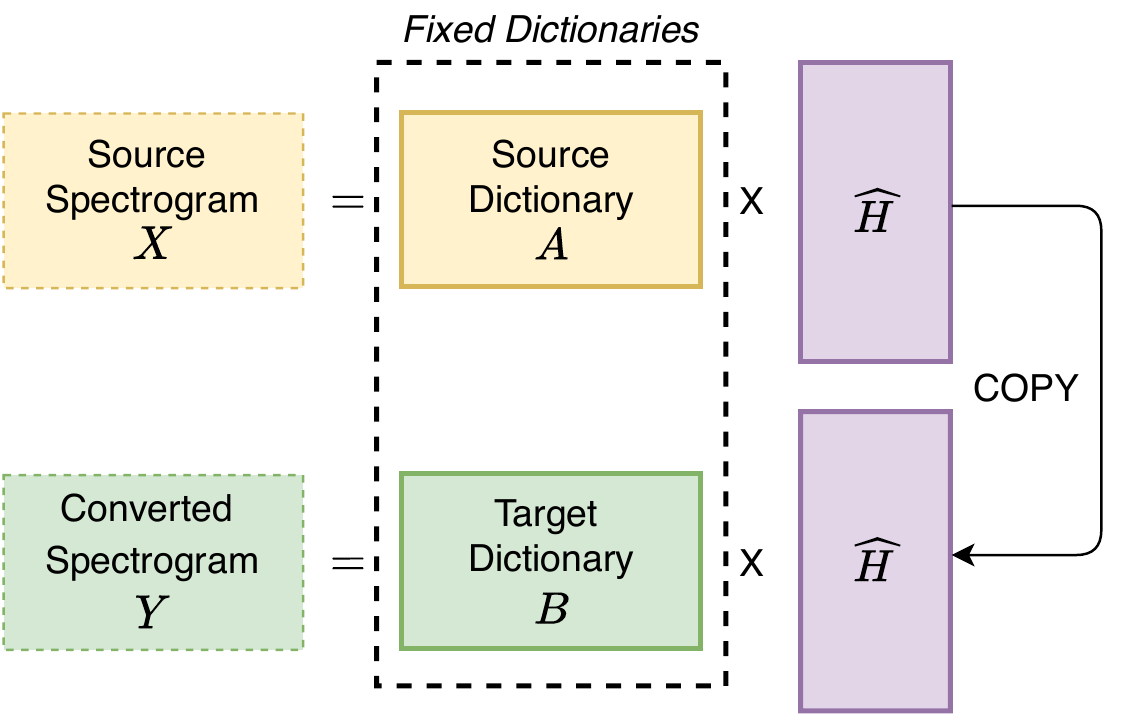}
 \caption{Illustration of non-negative matrix factorization for
exemplar-based sparse representation.}
 \label{nmf-section3}
\end{figure}
\subsubsection{Sparse Representation} 
One effective way to alleviate the over-smoothing problem is to apply sparsity constraint to the activation matrix, referred to as exemplar-based sparse representation.

As illustrated in Figure \ref{nmf-section3}, a pair of dictionaries $\bold{A}$ and $\bold{B}$ are first constructed from speech feature vectors, that we call aligned exemplars, from source and target. [$\bold{A}$; $\bold{B}$] is also called the coupled dictionary. At run-time, let's consider a speech utterance as a sequence of speech feature vectors, that form a spectrogram matrix. The matrix of a source utterance $\bold{X}$ can be represented as,
\begin{align}
\bold{X}\approx \bold{A}\bold{\hat{H}}
\end{align}
Due to the non-negative nature of spectrogram, NMF technique is employed to estimate the source activation matrix $\bold{\hat{H}}$, which is constrained to be sparse. Mathematically, we estimate $\bold{\hat{H}}$ by minimizing an objective function,
\begin{align}
\bold{\hat{H}}=\underset{\bold{H}\geq0}{\mathrm{argmin}}\text{ }d\left(\bold{X}, \bold{A}\bold{H}\right)+\lambda\bold{||H||}
\end{align}
where $\lambda$ is the sparsity penalty factor. To estimate activation matrix $\bold{\hat{H}}$, a generalised Kullback-Leibler (KL) divergence is used. It is assumed that source and target dictionaries $\bold{A}$ and $\bold{B}$ can share the same source activation matrix $\bold{\hat{H}}$. 

Therefore, the converted spectrogram for the target speaker can be written as,
\begin{align}
\label{eq:NMF_Y}
\bold{\hat{Y}}=\bold{B}\bold{\hat{H}}.
\end{align}
where the activation matrix $\bold{\hat{H}}$ serves as the pivot to transfer source utterance $\textbf{X}$ to target utterance $\textbf{Y}$.

The sparse representation framework continues to attract much attention in voice conversion. The recent studies include its extension to discriminative graph-embedded NMF approach \cite{Aihara2016}, phonetic sparse representation for spectrum conversion  \cite{ccicsman2017sparse}, and its application to timbre and prosody conversion\cite{Ming2016a, csicsman2017transformation}.

\subsubsection{Phonetic Sparse Representation}

As the frame-level mapping is done at
 acoustic feature level, the coupled dictionary [$\bold{A}$; $\bold{B}$] is therefore
called acoustic dictionary. With the scripts of the training data and a general purpose speech recognizer, we are able to obtain phonetic labels and their boundaries. Studies have shown that the strategy
of dictionary construction plays an
important role in voice conversion\cite{Kim2015}. The idea of selecting sub-dictionary according to the run-time speech content shows improved performance \cite{Aihara2014}. 

Phonetic sparse representation  \cite{ccicsman2017sparse} is an extension to sparse representation for voice conversion. It is built on the idea of phonetic sub-dictionaries, and dictionary selection at run-time. The study shows that  multiple phonetic sub-dictionaries consistently outperform single dictionary in exemplar-based sparse representation voice conversion \cite{ccicsman2017sparse, Aihara2014}. However, the phonetic sparse representation relies on a speech recognizer at run-time to help select the sub-dictionary.

\subsubsection{Group Sparse Representation}

Sisman et al. \cite{berrak-journal} proposed group sparse representation to formulate both exemplar-based sparse representation\cite{Wu2013}, and phonetic sparse representation \cite{ccicsman2017sparse} under a unified mathematical framework.  With the group sparsity regularization, only the phonetic
sub-dictionary that is relevant to the input features is
likely to be activated at run-time inference. Unlike phonetic sparse representation that relies on a speech recognizer for both training and run-time inference, group sparse representation only requires the speech recognizer during training when we build the phonetic dictionary. It was reported that group sparse representation provides similar performance to that of phonetic sparse representation when performing both spectrum and prosody conversion \cite{berrak-journal}.

\section{Statistical Modeling for Voice Conversion with Non-parallel Training Data}

It is easy to understand that it is more straightforward to train  a mapping function from parallel than non-parallel training data. However, parallel training data are not always available. In real-world applications, there are situations where only non-parallel data are available. Intuitively, if we can derive the equivalents of speech frames or segments between speakers from non-parallel data, we are able to establish or to refine the mapping function using the conventional linear transformation parameter training, such as GMM, DKPLS or frequency warping. 

There were a number of attempts to do so. For example,  one idea is to find source-target mapping between unsupervised feature clusters \cite{aligncluster}. Another is to use a speech recognizer to index the target training data so that we can retrieve similar frames from target database for a unknown source frame at run-time\cite {alignASR2004}. Unfortunately, each of the steps  may produce errors that accumulate and may lead to a poor parameter estimation\cite{aligncluster}.
There was also a study to use a hidden Markov model (HMM) that is trained for the target speaker, then the parameters of GMM-based
linear transformation function are estimated in such a way that the converted source vectors exhibit maximum likelihood with respect to the target HMM \cite{hmmnonpara2006}. This method shows comparable performance with methods of parallel data. However, it requires  that  the  orthography  of  the  training utterances  be  known, that limits its use.

Next we will discuss three clusters of studies and their representative work, 1) INCA algorithm,  2) unit selection algorithm, and 3) speaker modeling algorithm.

\begin{figure}[t]
  \centering
  \includegraphics[scale=0.65]{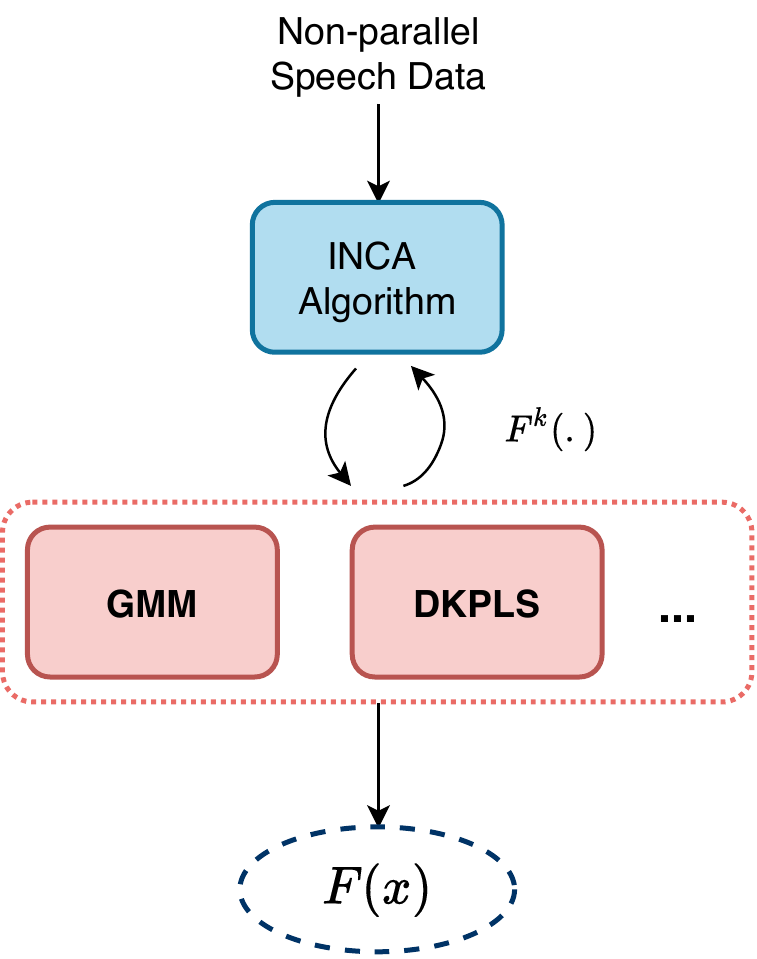}
 \caption{The training of a frame-wise mapping function is an iterative process between the nearest neighbor search step (INCA alignment) and the conversion step (a parametric mapping function).}
 \label{inca-section4}
\end{figure}
\subsection{INCA Algorithm}

INCA refers to an Iterative combination of a Nearest Neighbor
search step and a Conversion step Alignment method \cite{corsslingual-al2}. It learns a mapping function by finding the nearest neighbor of each source vector in the target acoustic space.
It is based on a hypothesis that an iterative refinement of the basic nearest neighbour method, in tandem with the voice conversion system, would lead to a progressive alignment improvement. The main idea is that the intermediate voice, $\textbf{x}^{k}_s$, obtained after the previous nearest neighbour alignment can be used as the source voice during the next iteration.
\begin{align}
    {\textbf{x}^{k+1}_s}={\mathsf{F^k}({\textbf{x}^k_s})}
\end{align}
During training, the optimization process is repeated until the current intermediate voice, $\textbf{x}^{k}_s$, is close enough to target voice, $\textbf{y}_t$. INCA represents a successful framework for the non-parallel training data problem, where the nearest neighbor search step (INCA alignment) and the conversion step (a parametric mapping function) iterates to optimize the mapping function, as illustrated in Figure \ref{inca-section4}.

INCA was first implemented with GMM approach \cite{stylianou1998continuous} for voice conversion to estimate a linear mapping function. As INCA does not
require any phonetic or linguistic information, it not only works for non-parallel training data, but also works for cross-lingual voice conversion. Experiments show that the INCA implementation of a cross-lingual system achieves similar performance to
its intra-lingual counterpart that is trained on parallel data \cite{corsslingual-al2}.

%\begin{figure*}
 % \centering
 % \includegraphics[scale=0.55]{Section5a.pdf}
 %\caption{Training and run-time conversion phases of a deep learning based %voice conversion system with parallel training data. The pink boxes represent %the training process, while the blue boxes represent the run-time process.}
 %\label{section5a.pdf}
%\end{figure*}

INCA was further implemented with DKPLS approach \cite{Helander2012} that was discussed in Section III.B for parallel training data. The idea \cite{Silen2012} is to use the INCA alignment algorithm \cite{corsslingual-al2} to find the corresponding frames from the source and target datasets, that allows the DKPLS regression to find a non-linear mapping between the aligned datasets. 
It was reported \cite{Silen2012} that the INCA-DKPLS implementation produces high-quality voice that is comparable to implementation with parallel training data on the same amount of training data.

\subsection{Unit Selection Algorithm}
%% add my PPG work with max 2 sentences
%% fujii2007highindividuality
  
Unit selection algorithms have been widely used to generate natural-sounding speech in speech synthesis. It is known to produce high speaker similarity and voice quality \cite{ stylianou2001applying, black1995optimising, fujii2007highindividuality} because the synthesized waveform is formed of sound units directly from the target speaker \cite{sagisaka1992atr}. The unit selection algorithm optimizes the unit selection from a voice inventory of a target speaker. It was suggested \cite{Erro2006VoiceCO} to make use of unit selection synthesis system to generate parallel versions of the training sentences from non-parallel data. With the resulting pseudo-parallel data, the statistical modeling techniques for parallel training data, that we discuss in Section III, can be readily applied. While this approach produces satisfactory voice quality \cite{Erro2006VoiceCO}, it requires a large speech database to develop the the voice inventory, that is not always practical in reality.

\begin{figure}[t]
  \centering
  \includegraphics[scale=0.7]{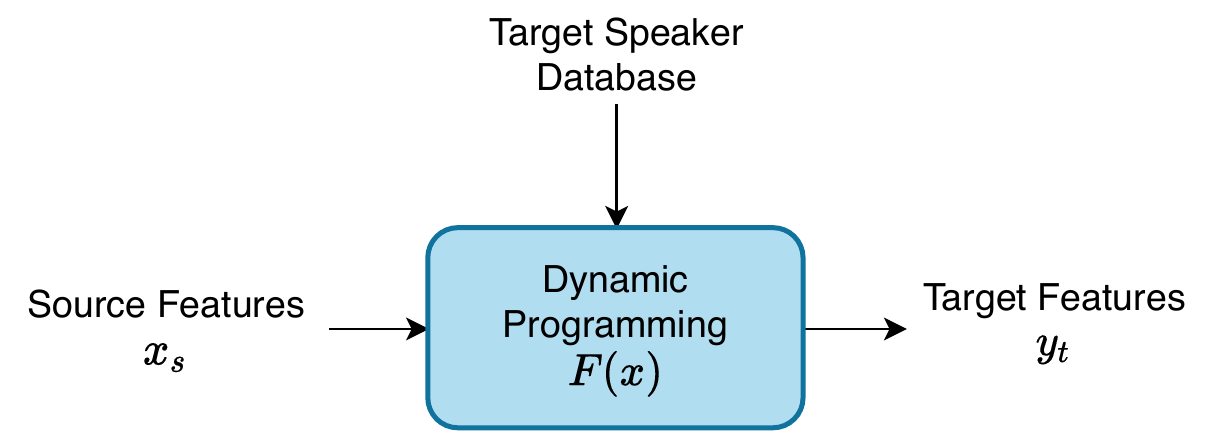}
 \caption{Run-time inference of unit selection algorithm that doesn't model a mapping function with parameters, but rather searches for output feature sequence directly from target speaker database, and optimizes the output at utterance level.}
 \label{unitselection-section4}
\end{figure}
Another idea is to follow what we do in unit selection speech synthesis by defining a speech feature vector as a unit \cite{sundermann2006text}. Given an utterance of $M$ speech feature vectors $\textbf{X} = \{ \textbf{x}_1, \textbf{x}_2, ..., \textbf{x}_M \}$ from the source speaker, a dynamic programming is applied to find the sequence of feature vectors $\textbf{y}_i$ from the target speaker, that minimizes a cost function,
\begin{align}
\label{eq:DTW}
%    \textbf{Y}&=\mathsf{F}(\textbf{X}) \\
    \textbf{Y}=\arg \min_{ y} \Big( \alpha \sum_{i=1}^{M} d_1(\textbf{x}_i,\textbf{y}_i) + (1-\alpha)  \sum_{i=2}^{M}d_2(\textbf{y}_i,\textbf{y}_{i-1}) \Big)
\end{align}
where $d_1(\cdot)$ represents the acoustic distance between a source and a target feature vector, while $d_2(\cdot)$ is the concatenative cost between two target feature vectors. With the acoustic distance, we make sure that the retrieved speech features from the target speakers are close to those of the source; with the concatenative cost, we encourage the consecutive speech frames from the target speaker database to be retrieved together in a multi-frame segment.
As illustrated in Figure \ref{unitselection-section4}, unit selection algorithm is a non-parametric solution because we don't model the conversion with parameters. It optimizes the output by applying a dynamic programming to find the best feature vector sequence from the target speaker database. The mapping function $\textbf{Y}=\mathsf{F}(\textbf{X})$ is defined by the cost function Eq.\ref{eq:DTW} itself, and optimized at the utterance level.

\subsection{Speaker Modeling Algorithm}

%The INCA and unit selection algorithms seek to find the best frame-level matching features between speakers. 
The techniques for text-independent speaker characterization are readily available for non-parallel training data, where a speaker can be modeled by a set of parameters, such as a GMM or i-vector. One is possible to make use such speaker models to perform voice conversion.

Mouchtaris et al.~\cite{Mouchtaris2006para-adapt} used a GMM-based technique to model relationship between reference speakers in advance and apply the relationship for a new speaker. Toda et al.  ~\cite{Toda2006EigenvoiceCB} proposed an  eigenvoice approach that performs two mappings,  one to map from the source speaker to an eigenvoice (or average voice) trained from reference speakers, and another from the eigenvoice to the target speaker. These approaches don't require parallel training data, they do require parallel data from some reference speakers. 

In speaker verification, the joint factor analysis method \cite{dehak2010front} decomposes a supervector into speaker independent, speaker dependent and channel dependent components, each of which is represented by a low-dimensional set of factors. This aims to disentangle speaker from other speech content for effective speaker verification.  Inspired by this idea, we argue~\cite{Zhizheng2012FactorAnal} that similar decomposition would be useful in voice conversion, where we would like to separate speaker information from the linguistic content, and apply factor analysis on the speaker specific component.

With factor analysis, the speaker specific component can be represented by a low-dimensional set of latent variables via the factor loadings. One of the ideas~\cite{Zhizheng2012FactorAnal} is to estimate the phonetic component and factor loadings from non-parallel prior data. In this way, during the training process, we only estimate a low-dimensional set of speaker identity factors and a tied covariance matrix instead of a full conversion function from the source-target parallel utterances. Even though parallel utterances are still required for estimating the conversion function, the use of prior data allows us to obtain a reliable model from much fewer training samples than those required by conventional JD-GMM \cite{Stylianou1998}.

Another idea is to perform the voice conversion in i-vector \cite{dehak2010front} speaker space, where i-vector is used to disentangle a speaker from the linguistic content. The primary motivation is that an i-vector can be extracted in an unsupervised manner regardless of speaker or speech content, which
opens up new possibilities especially for non-parallel data scenarios
where source and target speech is of different content or even in different languages\cite{Sun2016, zhouyi, corsslingual-al1}.
  Kinnunen et al.
\cite{kinnunen2017non} study a way to shift the acoustic features of input speech towards target speech in the i-vector space. The idea is to learn a function that maps the i-vector of the source utterance to that of the target. With the mapping function, we are able to convert the source speech frame-by-frame to the target. This technique is free of any parallel data, and text transcription. 

\section{Deep Learning for Voice Conversion}

Voice conversion is typically a research problem with scarce training data. Deep learning techniques are typically data driven, that rely on big data. However, this is actually the strength of deep learning in voice conversion. Deep learning opens up many possibilities to benefit from abundantly available training data, so that the voice conversion task can focus more on learning the mapping of speaker characteristics. For example, it shouldn't be the job of voice conversion task to infer low level detail during speech reconstruction, a neural vocoder can learn from large database to do so~\cite{wavenet-vae2019}. It shouldn't be a task of voice conversion to learn how to represent an entire phonetic system of a spoken language, a general purpose acoustic model of neural ASR~\cite{miyoshi2017voice} or TTS~\cite{cotatron} system can learn from a large database to do so. By leveraging the large database, we free up the conversion network from using its
capacity to represent low level detail and general information, but instead, to focus on the high level semantics necessary  for speaker identity conversion.  

Deep learning techniques also transform the way we implement the analysis-mapping-reconstruction pipeline. For effective mapping, we need to derive adequate intermediate representation of speech, that was discussed in Section II. The concept of \textit{embedding} in deep learning provides a new way of deriving the intermediate representation, for example, latent code for linguistic content, and speaker embedding for speaker identity. It also makes the disentanglement of speaker from speech content much easier.

%In the context of voice conversion, speech carries both speaker-dependent information and speaker-independent information.  

In this section, we will summarize how deep learning helps address existing research problems, such as parallel and non-parallel data voice conversion. We will also review how deep learning breaks new ground in voice conversion research.

\subsection{Deep Learning for Frame-Aligned Parallel Data}

The study on deep learning approaches for voice conversion started with parallel training data, where we use a neural network as an improved regression function to approximate the frame-wise mapping function $\textbf{y}=\mathsf{F}({\textbf{x}})$ under the frame-level mapping paradigm in Figure \ref{statistical-withparallel}. 

\subsubsection{DNN Mapping Function}

The early studies on DNN-based voice conversion methods are focused on spectral transformation. DNN mapping function, $\textbf{y}=\mathsf{F}({\textbf{x}})$, has some clear advantage over other statistical models, such as GMM, and DKPLS. For instance, it allows for non-linear mapping between source
and target features, and there is little restriction to the dimension of features to be modeled. We note that conversion on other acoustic features, such as fundamental frequency and energy contour, can also be done similarly \cite{xie2014pitch}. 

Desai et al.  \cite{desai2009voice} proposed a DNN to map a low-dimensional spectral representation, such as mel-cepstral coefficients (MCEP), from source to target speaker. Nakashika et al.  \cite{nakashika2013voice} proposed to use  Deep Belief Nets (DBNs) to extract latent features from source and target cepstrum coefficients, and use a neural network with one hidden layer to perform conversion between latent features. Mohammadi et al. \cite{mohammadi2014voice} furthered the idea by studying a deep autoencoder from multiple speakers to derive a compact representations of speech spectral feature. High-dimensional representation of spectrum has also been used in a more recent work \cite{xie2014sequence} for spectral mapping, together with dynamic features and a parameter generation algorithm \cite{tokuda2000speech}. Chen et al. \cite{Chen2014} proposed to model the distributions of spectral envelopes of source and target speakers respectively through a layer-wise generative training. 

Generally speaking, DNN for spectrum and/or prosody transformation requires a large amount of parallel training data from paired speakers, which is not always feasible. But it opens up opportunities for us to make use of speech data from multiple speakers beyond source and target, to better model the source and the target speakers, and to discover better feature representations for feature mapping. 

\subsubsection{LSTM Mapping Function}
To model the temporal correlation across speech frames in voice conversion, Nakashika et al. \cite{Nakashika2014} explore the use of Recurrent Temporal Restricted Boltzmann Machines (RTRBM), a type of recurrent neural networks. The success of Long-Short Term Memory (LSTM) \cite{hochreiter1997long, gers1999learning} in sequence to sequence modeling inspires the study of LSTM in voice conversion, which leads to an improvement of naturalness and continuity of the speech output.

The LSTM network architecture consists of a set of memory blocks and gates, that support the storage and access to long-range contextual information \cite{greff2016lstm}. LSTM can learn the optimal amount of contextual information for voice conversion. A bidirectional LSTM (BLSTM) network is expected to capture sequential information and maintain long-range contextual features from both forward sequence and backward sequence \cite{Sun2016}. 

Sun et al.~\cite{Sun2015} and Ming et al.~\cite{Ming2016} proposed a deep bidirectional LSTM network (DBLSTM) by stacking multiple hidden layers of BLSTM network architecture, that is shown to outperform DNN voice conversion even without using dynamic features.
%\textcolor{red}{not to include here. Some successful implementations include DBLSTM-RNNs \cite{emir2019semantically, ak2020semantically}}.
While DBLSTM-based voice conversion approach generates high-quality synthesized voice, it typically requires a large speech corpus from source and target speakers for training, that limits the scope of the applications in practice \cite{Sun2015}.

Just like GMM approach, DNN and LSTM techniques %above are frame-based method that maps speech feature vectors frame-by-frame, hence does not capture the temporal dependencies of speech sequences \cite{Sun2015}. They
rely on external frame aligner during training data preparation, as illustrated in Figure  \ref{statistical-withparallel}. At run-time, the conversion process follows the typical flow of 3-step pipeline, and doesn't change the speech duration during the conversion.

\subsection{Encoder-decoder with Attention for Parallel Data}
\begin{figure}[t]
  \centering
  \includegraphics[scale=0.6]{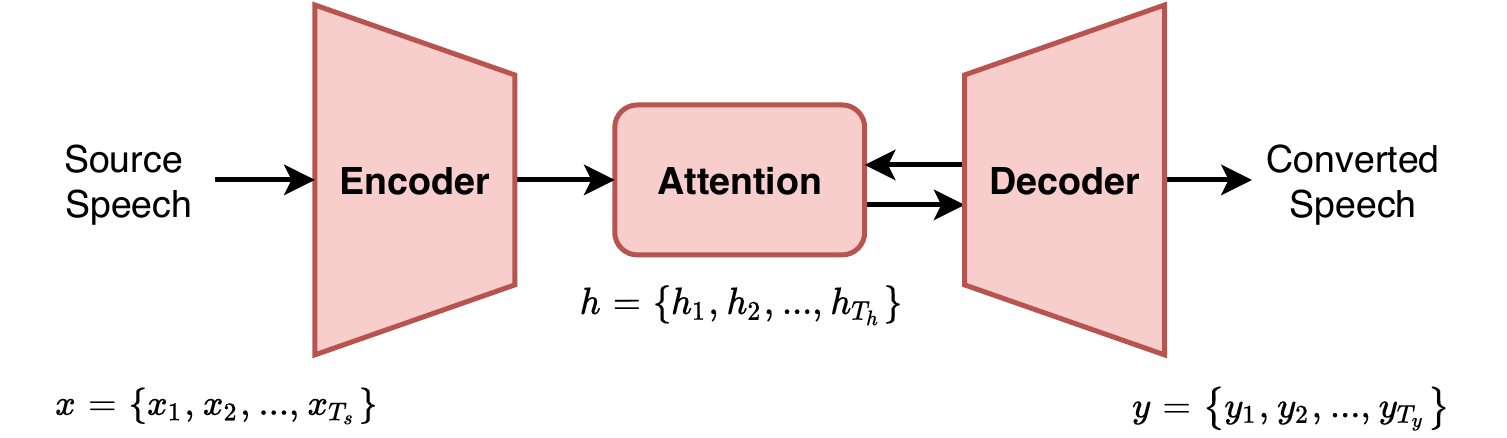}
 \caption{Encoder-decoder mechanism with attention for voice conversion. }
 \label{section5b-encoderdecoder}
\end{figure}

The research problems of voice conversion are centered around alignment and mapping, which are interrelated both during training and at run-time inference, as illustrated in Figure 2.  During training, more accurate alignment helps build better mapping function, that explains why we prefer parallel training data. At run-time inference, the frame-level mapping paradigm doesn't change the duration of the speech during the conversion. While it is possible to model and predict the duration for voice conversion output, it is not straightforward to incorporate duration model and mapping model in a systematic manner. Deep learning provides a new solution to this research problem.

The attention mechanism \cite{attentionMT,Attention2017} in encoder-decoder structure neural network brings about a paradigm change. The idea of attention was first successfully used in machine translation \cite{attentionMT}, speech recognition \cite{attentionASR}, and  sequence-to-sequence speech synthesis \cite{wang2017tacotron, tacotron2, ping2017deep, tachibana2018efficiently}, that led to many parallel studies in voice conversion \cite{zhang2019sequence, attS2S2019}. 
With the attention mechanism, the neural network learns the feature mapping and alignment at the same time during training. At run-time inference, the network automatically decides the output duration according to what it has learnt. In other words, the frame-aligner in Figure  \ref{statistical-withparallel} is no longer required.

\begin{figure*}
  \centering
  \includegraphics[scale=0.52]{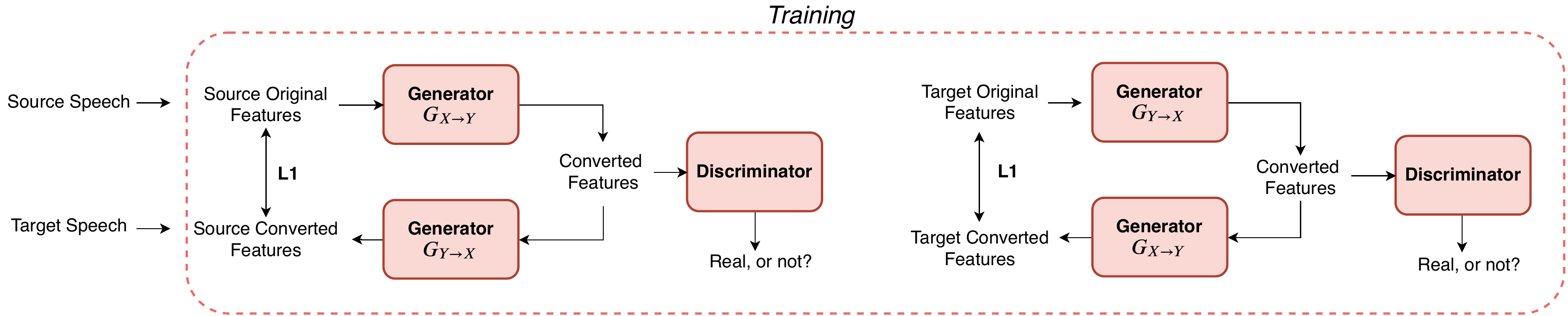}
 \caption{Training a CycleGAN with cycle-consistency loss of L1 norm for voice conversion with non-parallel training data of paired speakers. L1 norm represents the least absolute errors}
 \label{section5b1.pdf}
\end{figure*}

There are several variations based on recurrent neural networks, such as sequence-to-sequence conversion network (SCENT) \cite{zhang2019sequence}, and AttS2S-VC \cite{attS2S2019}. They follow the widely-used architecture of
encoder-decoder with attention \cite{cho2014learning, luong2015effective}.  Suppose that we have a source speech $\textbf{x}=\{\textbf{x}_1, \textbf{x}_2, ..., \textbf{x}_{T_s} \}$. The encoder network
first transforms the input feature sequences into hidden representations, $\textbf{h}=\{\textbf{h}_1, \textbf{h}_2, ..., \textbf{h}_{T_h} \}$ at a lower frame rate with $T_h < T_s$, which are suitable for the decoder to deal with. At each decoder time step, the attention module aggregates the encoder outputs by attention probabilities and produces
a context vector. Then, the decoder predicts output acoustic
features frame by frame using context vectors. Furthermore, a
post-filtering network is designed to enhance the accuracy of
the converted acoustic features to generate the converted speech $\textbf{y}=\{\textbf{y}_1, \textbf{y}_2, ..., \textbf{y}_{T_y} \}$. During training, the attention mechanism learns the mapping dynamics between source sequence and target sequence. At run-time inference, the decoder and the attention mechanism interacts to perform the mapping and alignment at the same time.
 The overall architecture is illustrated in Figure \ref{section5b-encoderdecoder}.
 
 While recurrent neural networks represent an effective implementation for sequence-to-sequence conversion, recent studies have shown that convolutional neural networks also learn well the long-term dependencies\cite{van2016wavenet, Gehring2017ConvolutionalST}. It employs an attention mechanism that effectively makes possible parallel computations for encoding and decoding. During decoding, the causal convolution design allows the model to generate an output sequence in an autoregressive manner. Kameoka et al. proposed a convolutional neural networks implementation for voice conversion \cite{Kameoka2018ConvS2SVCFC}, that is called ConvS2S-VC.  Recent studies show that ConvS2S-VC outperforms its recurrent neural network counterparts in both pairwise and many-to-many voice conversion \cite{attS2S2019}.   
  
The encoder-decoder structure with attention marks a departure from the frame-level mapping paradigm. The attention doesn't perform the mapping frame-by-frame, but rather allows the decoder to attend to multiple speech frames and uses the soft combination to predict an output frame in the decoding process. With the attention mechanism, the duration of the converted speech $T_y$ is typically  different from that of the source speech $T_s $ to reflect the differences of speaking style between source and target. This represents a way to handle both spectral and prosody conversion at the same time.  The studies have  attributed the improvement of voice quality to the effective attention mechanism. The attention mechanism also represents the first step towards relaxing the rigid requirement of parallel data in voice conversion.

\subsection{Beyond Parallel Data of Paired Speakers}

In Section III and IV, we study statistical modeling for voice conversion with parallel training data and non-parallel training data. The advent of deep learning has broken new ground for voice conversion research. We now go beyond the paradigm of parallel and non-parallel training data.  Traditionally, \textit{nonparallel training data} refers to the case where nonparallel utterances from source-target speaker pair are required. However, the recent studies show that, deep learning has enabled many voice conversion scenarios without the need of parallel data. In this section, we summarize the studies into four scenarios,

\begin{enumerate}
    \item Non-parallel data of paired speakers,
    \item Leveraging TTS systems,
    \item Leveraging ASR systems, and
    \item Disentangling speaker from linguistic content.
%    \item Using external ASR for unseen source and/or target speakers, and
%    \item Disentangling speaker and/or content representations for unseen source and/or target speakers.
\end{enumerate}

\subsubsection{Non-parallel data of paired speakers}

Voice conversion with non-parallel training data is a task similar to image-to-image translation \cite{isola2017image, zhu2017unpaired, choi2018stargan, liu2019few, ak2018learning}, which is to find a mapping from a source domain to a target domain without the need of parallel training data. 
%Image-to-image translation is to learn the mapping between an input image and an output image using a training set of aligned image pairs. 
%It is considered that the task of converting the speaking voice from source to target while preserving the linguistic content is similar to that of translating 
Let's draw a parallel between image-to-image translation and voice conversion. In image translation, we would like to translate a horse to a zebra, where we preserve the structure of horse and change the coat of horse to that of zebra \cite{image1, ak2019attribute, ak2019deep, ak2018efficient, choi2020stargan}, in voice conversion, we would like to transform one voice to that of another, while preserving the linguistic, and prosodic content. 
%both in the absence of parallel training examples. Inspired by the results in image processing, there have been several successful CycleGAN implementation in voice conversion \cite{kaneko2017parallel,kaneko2018cyclegan,kaneko2019cyclegan, cycleGAN2}.

CycleGAN is based on the concept of adversarial learning \cite{goodfellow2014generative}, which is to train a generative model to find a solution in a min-max game between two neural networks, called as generator ($G$) and discriminator ($D$). It is known to achieve remarkable results \cite{image1} on several tasks where paired training data does not exist, such as image manipulation and synthesis \cite{image1, image2, image3, ak2019deep, ak2020semantically, emir2019semantically, wang2018high}, speech enhancement \cite{cycleGAN_speechen}, speech recognition \cite{cycleGAN_ASR}, speech synthesis \cite{yook2018voice, jia2020speech}, and music translation \cite{huang2018timbretron}.
 
 As the speech data are non-parallel, alignment is not easily achieved. Kaneko and Kameoka first studied a CycleGAN \cite{kaneko2017parallel,kaneko2018cyclegan,kaneko2019cyclegan, cycleGAN2} that incorporates three loss functions: adversarial loss, cycle-consistency loss, and identity-mapping loss, to learn forward and inverse mapping between source and target speakers.

The adversarial loss measures how distinguishable between the data distribution of converted features and source features $\textbf{x}$ or target features $\textbf{y}$. For the forward mapping, it is defined as follows:
\begin{multline}
   L_{ADV}(G_{X\rightarrow Y},D_Y,X,Y) = \mathbb{E}_{y\sim P(y)}[D_Y(\textbf{y})] \\
    +\mathbb{E}_{x\sim P(x)}[log(1-D_Y(G_{X\rightarrow Y}(\textbf{x}))]
\end{multline}
The closer the distribution of converted data with that of target data, the smaller the loss becomes. 

The adversarial loss only tells us whether $G_{X\rightarrow Y}$ follows the distribution of target data and does not ensure that the contextual information, that represents the general sentence structure we would like to carry over from source to target, is preserved. To ensure that we maintain the consistent contextual information between $\textbf{x}$ and $G_{X\rightarrow Y} (\textbf{x})$, the cycle-consistency loss, that is presented in Figure \ref{section5b1.pdf}, is introduced,
\begin{multline}
    L_{CYC}(G_{X\rightarrow Y},G_{Y\rightarrow X}) \\
    = \mathbb{E}_{\textbf{x}\sim P(\textbf{x})}[\Vert G_{Y\rightarrow X}(G_{X\rightarrow Y}(\textbf{x}))-\textbf{x}\Vert_1]\\
    +\mathbb{E}_{\textbf{y}\sim P(y)}[\Vert G_{X\rightarrow Y}(G_{Y\rightarrow X}(\textbf{y}))-\textbf{y}\Vert_1]
\end{multline}
{where $\Vert \: \cdot \: \Vert_1$ refers to a L1 norm function, or least absolute errors, that is known to produce sharper spectral features.} This loss encourages $G_{X\rightarrow Y}$ and $G_{Y\rightarrow X}$ to find an optimal pseudo pair of $(\textbf{x},\textbf{y})$ through circular conversion. 

To encourage the generator to find the mapping that preserves underlying linguistic content between the input and output \cite{Taigman2016UnsupervisedCI},  an identity mapping loss is introduced as follows,
\begin{multline}
    L_{ID}(G_{X\rightarrow Y},G_{Y\rightarrow X}) \\
    = \mathbb{E}_{\textbf{x}\sim P(x)}[\Vert G_{Y\rightarrow X}(\textbf{x})-\textbf{x}\Vert]
    + \mathbb{E}_{\textbf{y}\sim P(y)}[\Vert G_{X\rightarrow Y}(\textbf{y})-\textbf{y}\Vert]
\end{multline}
Combining these three loss functions, we can obtain the overall loss function of CycleGAN \cite{kaneko2017parallel,kaneko2018cyclegan}. 
%\begin{multline}
 %   L(G,F,D_X,D_Y,X,Y) = L_{GAN}(G,D_Y,X,Y)\\
  %  +L_{GAN}(F,D_X,X,Y)+\lambda_{CYC}L_{CYC}(G,F,X,Y)\\
   % +\lambda_{ID}L_{ID}(G,F,X,Y)
%\end{multline}
%where $\lambda_{CYC}$ and $\lambda_{ID}$ are trade-off parameters. 

%The optimal mapping functions $G^*$ and $F^*$ are obtained by solving the minmax-game defined as:
%\begin{equation}
 %  G^*,F^*=\mathop{\arg\min}_{G,F}\mathop{\max}_{D_X,D_Y}L(G,F,D_X,D_Y,X,Y)
%\end{equation}

%\textcolor{red}{Speech is known to have sequential and hierarchical structures, e.g., voiced or unvoiced segments and phonemes or morphemes. An effective way to represent such structures would be to use an RNN, but it is computationally demanding due to the difficulty of parallel implementations. Instead, CycleGAN with gated CNNs that not only allow parallelization over sequential data but also achieve state-of-the-art in language modeling \cite{dauphin2017language}, speech modeling \cite{kaneko2017sequence} as well as voice conversion \cite{tobing2019voice}.}

CycleGAN represents a successful deep learning implementation to find an optimal pseudo pair from non-parallel data of paired speakers. It doesn't require any frame alignment mechanism such as dynamic time warping or attention.  Experimental results show that, with non-parallel training data, CycleGAN achieves comparable performance to that of GMM-based system that is trained on twice amount of parallel data \cite{kaneko2017parallel}. Moreover, with the adversarial training, it effectively overcomes the over-smoothing problem, which is known to be one of the main factors leading to speech-quality degradation. We note that more recently, CycleGAN-VC2, an improved version of CycleGAN-VC has been studied \cite{kaneko2019cyclegan}, that further improves CycleGAN by incorporating three new techniques: an improved objective (two-step adversarial losses), improved generator (2-1-2D CNN), and improved discriminator (PatchGAN). CycleGAN has been successfully applied in mono-lingual \cite{cycleGAN2, tobing2019voice},  cross-lingual voice conversion \cite{sismanstudy}, emotional voice conversion \cite{zhou2020transforming, zhou2020converting} and rhythm-flexible voice conversion \cite{yeh2018rhythm}.

Unlike the encoder-decoder structure, CycleGAN follows a generative modeling architecture that doesn't explicitly model some internal representations to support flexible manipulation, such as voice identity, duration of speech, and emotion. Therefore, it is more suitable for voice conversion between a specific source and target pair. Nonetheless, it represents an important milestone towards non-parallel data voice conversion.

\subsubsection{Leveraging TTS systems}
\begin{figure}[t]
  \centering
  \includegraphics[scale=0.62]{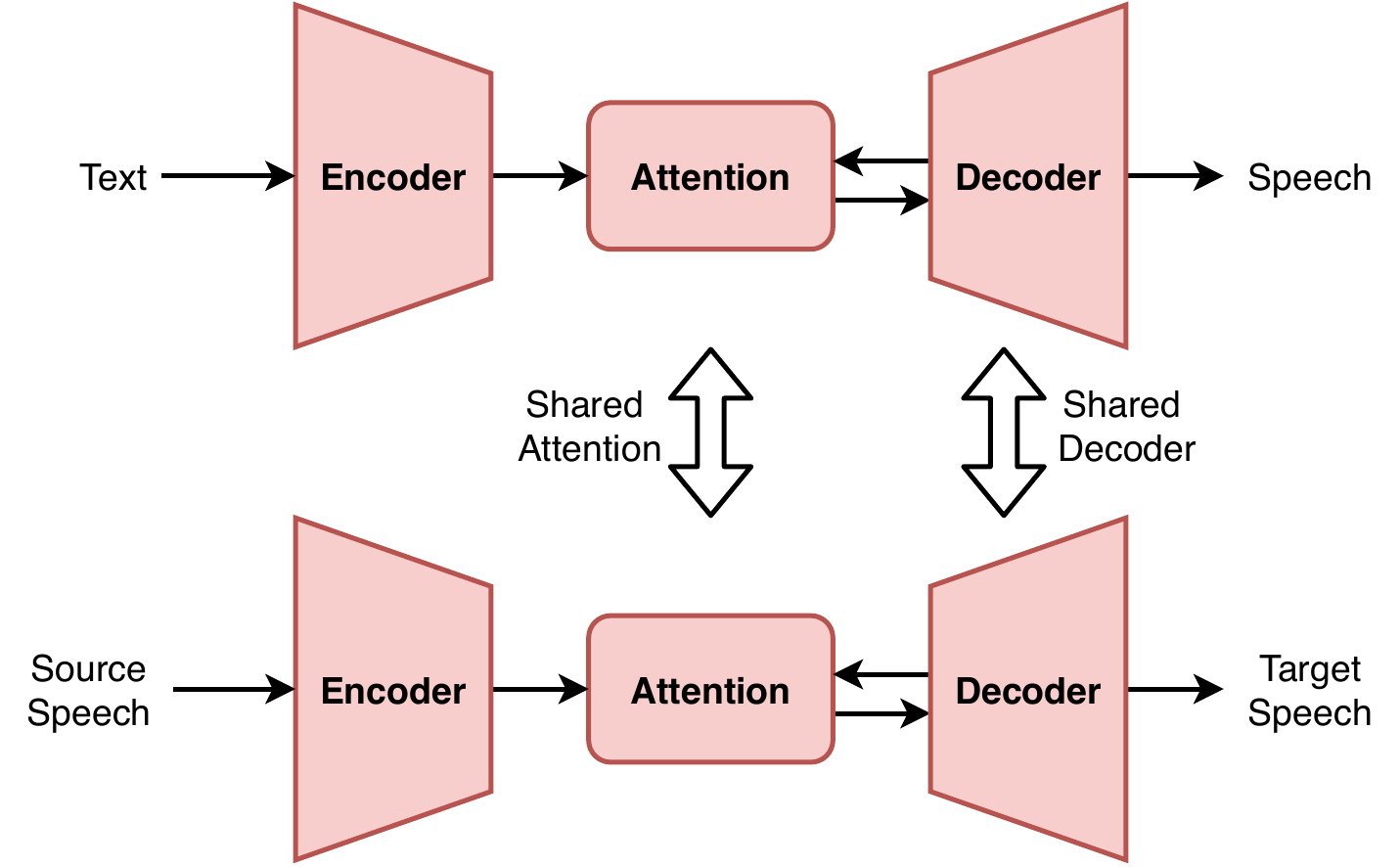}
 \caption{The upper panel is a TTS flow, and the lower panel is a voice conversion flow. Both follow similar encoder-decoder with attention architecture. The voice conversion network leverages the TTS network, that is linguistically informed. }
 \label{section5c-tts}
\end{figure}
%The typical flow of voice conversion in Figure \ref{flow.pdf} takes a source voice as input and generates the target voice as output. 
We have discussed the deep learning architectures for voice conversion that do not involve text.
One of the important aspects of voice conversion is to carry forward the linguistic content from source to target. Voice conversion and TTS systems are similar in the sense that they both aim to generate high quality speech with the appropriate linguistic content. A TTS system provides a mechanism for the speech to adhere to the linguistic content. 
The ideas to leverage TTS mechanism can be motivated in different ways. Firstly, a TTS system is trained on a large speech database that offers a high quality speech re-construction mechanism given the linguistic content; secondly,  a TTS system is equipped with a quality attention mechanism that is needed by voice conversion. 

Encoder-decoder models with attention have recently shown considerable success in modeling a variety of complex sequence-to-sequence problems. Tacotron \cite{wang2017tacotron, liu2019teacher, liu2020wavetts, liu2020modeling, liu2020expressive} represents one of the successful text-to-speech (TTS) implementations, that has been extended to voice conversion\cite{zhang2019sequence, zhang2019joint,zhang2020transfer}. The strategy to leverage TTS knowledge is built on the ideas of shared attention knowledge and/or shared decoder architecture as illustrated in Figure \ref{section5c-tts}. Zhang et al.~\cite{zhang2020transfer} proposed a transfer learning technique for voice conversion network to learn from the phonetic context vectors derived from a TTS attention mechanism, and to share the decoder with a TTS system, that represents a typical example of such leverage.   %\textcolor{red}{that leverage 1) joint training of TTS and VC (Mingyang + Yamagishi), 2) voice conversion supervised and/or improved by TTS (supervised - Seoul Nat. Univ/ improved - Paratron); } %3) voice conversion with TTS as a teacher.

%and automatic speech recognition (ASR) \cite{chiu2018state}, using a single neural network that directly generates the target sequences.}

%In voice conversion, sequence-to-sequence modeling has shown to achieve remarkable performance with appropriate duration conversion \cite{zhang2019sequence}. Moreover, with text supervision, it is possible to improve the performance even further \cite{zhang2019improving}. 
Zhang et al. proposed a joint training system architecture for both text-to-speech and voice conversion~\cite{zhang2019joint} by extending the model architecture of Tacotron, which features a multi-source sequence-to-sequence model with a dual input, and dual attention mechanism.  By taking only text as input, the system performs speech synthesis. The system can also take either voice alone, or both text and voice, denoted as hybrid TTS \& VC, as input for voice conversion. The multi-source encoder-decoder model is trained with a decoder that is linguistically informed via the TTS joint training, as illustrated as {shared decoder} in Figure \ref{section5c-tts}. Experiments show that the joint training has improved the voice conversion task with or without text input at run-time inference. 

Park et al. proposed a voice conversion system, known as Cotatron, that is built on top of a multi-speaker Tacotron TTS architecture~\cite{cotatron}. At run-time inference, the pre-trained TTS system is used to derive speaker-independent linguistic features of the source speech. This process is guided by the transcription of the input speech, as such, text transcription of source speech is required at run-time inference. The system uses the TTS encoder to extract speaker-independent linguistic features, or disentangle the speaker identity. The decoder then takes the attention-aligned speaker-independent linguistic features as the input, and the target speaker identity as the condition, to generate a target speaker's voice. In this way, voice conversion  leverage the attention mechanism or \textit{shared attention} from TTS, as shown in Figure \ref{section5c-tts}.  Cotatron is designed to perform  one-to-many voice conversion. A study \cite{huang2019voice}, that shares similar motivation with \cite{cotatron} but is based on the Transformer instead of Tacotron,  suggests transferring  knowledge from a learned TTS model to benefit  from large-scale, easily accessible TTS corpora. 

%\textcolor{blue}{Recently, an end-to-end speech-to-speech sequence transducer Parrotron \cite{biadsy2019parrotron} that combines attention-based speech recognition and synthesis models, has been proposed. In this model, speech spectrogram is generated as a function of a different input spectrogram, with no intermediate discrete representation. It is reported that pre-trained normalization model of Parrotron can be adapted to perform a more challenging task of converting highly atypical speech from a deaf speaker into fluent speech, that significantly improving intelligibility and naturalness. We note that Parrotron training requires a large corpus that consists of utterances from many speakers. Since it is impractical to have single speaker record many hours of speech in clean acoustic environment, a WaveNet-based TTS is used to generate parallel corpus to eliminate noise and disfluencies. With the help of a TTS model, we can synthesize as much data as necessary to scale to very large corpora. Parrotron shows that, given sufficient training data, an end-to-end trained one-to-one conversion model can dramatically improve intelligibility and naturalness of a deaf speaker.}

Zhang et al. \cite{zhang2019improving} proposed to improve the sequence-to-sequence model\cite{zhang2019sequence} by using text supervision during training. A multi-task learning structure is designed which adds auxiliary classifiers to the middle layers of the sequence-to-sequence  model to predict linguistic labels as a secondary task. The linguistic labels can be obtained either manually or automatically with alignment tools. With the linguistic label objective, the encoder and decoder are expected to generate meaningful intermediate representations which are linguistically informed. The text transcripts are only required during training.  Experiments show that the multi-task learning with linguistic labels effectively improves the alignment quality of the model, thus alleviates issues such as mispronunciation.

\begin{figure}[t]
  \centering
  \includegraphics[scale=0.6]{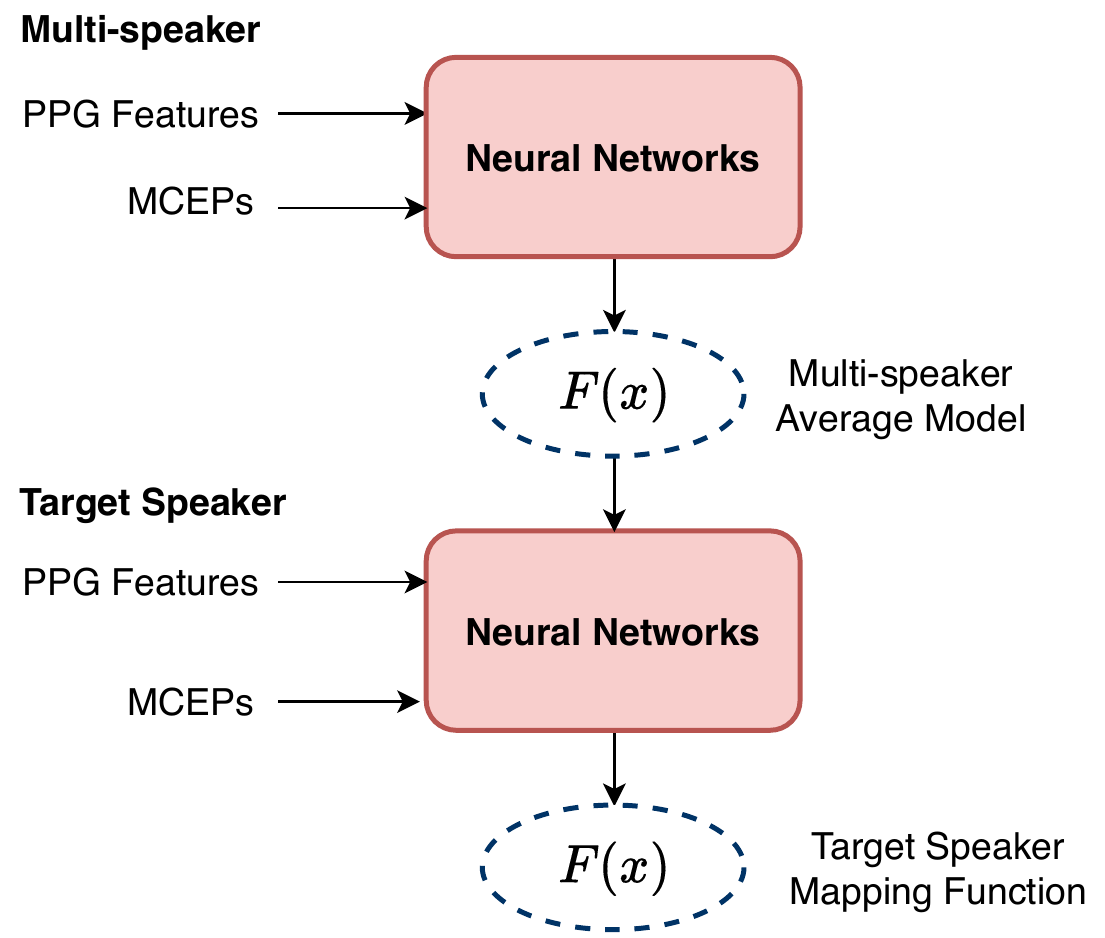}
 \caption{Training phase of the average modeling approach that maps PPG features to MCEP features for voice conversion~\cite{Tianavmodel}.}
 \label{section5b3.pdf}
\end{figure}

The neural representation of deep learning has facilitated the interaction between TTS and voice conversion. By leveraging TTS systems, we hope to improve the training and run-time inference of voice conversion with by adhering to linguistic content. However, such techniques usually require a large training corpus. Recent studies introduced a framework for creating limited-data VC system \cite{luong2019bootstrapping,huang2019voice,luong2020nautilus} by bootstrapping from a speaker-adaptive TTS model. It deserves future studies as to how voice conversion can benefit from TTS systems without involving large training data.

%\subsubsection{Improving VC using external ASR for unseen source and/or target speakers}
%Most  of  the  voice  conversion  techniques  rely  on  parallel data  during  training. When parallel  data  are  not available, there have been attempts to find source-target paired frames from non-parallel corpus \cite{cycle_gan, Hsu2017_gan}. 

%% THE PAPER PROF LI SENT

\subsubsection{Leveraging ASR systems}

Deep learning approaches for voice conversion typically require a large  parallel corpus for training. This is partly because we would like to learn the latent representations that describe the phonetic systems. The requirement of training data has limited the scope of potential applications. We know that most ASR systems are already trained with a large corpus. They already describe well the phonetic systems in different ways. The question is how to leverage the latent representations in ASR systems for voice conversion.

%one challenge in voice conversion would be how best to train voice conversion model using a limited amount of training data from source and/or target speakers, while leveraging on publicly available large corpus and/or tools such as ASR. 

One of the ideas is to use the context posterior probability sequence produced by the ASR model with sequence to sequence learning to generate a target speech feature sequence\cite{miyoshi2017voice}. In this model, the system has an encoder-decoder structure similar to Figure 6, except that it uses a speech recognizer as the encoder, and a speech synthesizer as the decoder. Another study is to guide a sequence to sequence voice conversion model by an ASR system, which augments inputs with bottleneck features \cite{zhang2019sequence}. Recently, an end-to-end speech-to-speech sequence transducer, Parrotron \cite{biadsy2019parrotron}, was studied. Parrotron learns to convert speech spectrogram of any speakers, with multiple accents and imperfections, to the voice of a single predefined target speaker. Parrotron accomplishes this by using an auxiliary ASR decoder to predict the transcript of the output speech, conditioned on the encoder latent representation. The multi-task training of Parrotron optimizes the decoder to generate the target voice, at the same time, constrains the latent representation to retain linguistic information only. The ASR decoder aims to disentangle the speaker’s identity from the speech. The above techniques adopt the encoder-decoder with attention architecture.

It is another way to look at voice conversion that speech consists of two components, speaker dependent component and speaker independent component. If we are able to decompose speech signals into the two components, we can carry over the former, and only convert the latter to achieve voice conversion. The average modeling technique represents one  of the successful implementations \cite{Wu}, where we build a mapping function to convert phonetic posteriogram (PPG)~\cite{lifasun} to acoustic features. The PPG features are derived from an ASR system, that can be considered as speaker independent. We train the mapping function from multi-speaker, non-parallel speech data. In this way, one doesn't need to train a full conversion model for each target speaker. The average model can be adapted towards the target with a small amount of target speech. The training and adaptation of the average model are illustrated in Figure \ref{section5b3.pdf}. 

%The average model learns the mapping between PPG linguistic features and acoustic features of the same speaker. Therefore, parallel training data is not required. 
There were several follow-up studies along this direction, for example, Tian et al. propose a PPG to waveform conversion~\cite{tian2019speaker}, and a average model with speaker identity~\cite{dehak2010front} as a condition~\cite{Tianavmodel}. Zhou et al. propose to use PPG as the linguistic features for cross-lingual voice conversion~\cite{zhouyi}. Liu et al. propose to use PPG for emotional voice conversion \cite{liu2020multi}. Zhang et al. also show that the average model framework can benefit from a small amount of parallel training data using an error reduction network \cite{zhang2018error}.  

%\textcolor{red}{It has found applications in mono-lingual  voice conversion \cite{Tianavmodel, zhang2018error}, cross-lingual voice conversion \cite{zhouyi} and emotional voice conversion \cite{liu2020multi}.}

%it is considered that  linguistic features  are  speaker  independent  that  describe the phonetic content of the utterance, while acoustic features are speaker dependent. Therefore, it is possible to learn a linguistic-acoustic mapping from the same utterance.  In other words, we do not need any parallel data in either average model training or adaptation. 

%Moreover, together with i-vectors \cite{garcia2011analysis, senior2014improving}, a compact vector representing the identities of source and target speakers, personalized target voice can be generated, both for seen and unseen speakers. Subjective evaluation has confirmed the effectiveness of the proposed approach.

\subsubsection{Disentangling speaker from linguistic content}
%for unseen source and/or target speakers}

In the context of voice conversion, speech can be considered as a composition of speaker voice identity and linguistic content. If we are able to disentangle speaker from the linguistic content, we can change the speaker identity independently of the linguistic content. Auto-encoder\cite{vae_gan} represents one of the common techniques for speech disentanglement, and reconstruction. There are other techniques such as instance normalization~\cite{Chou2019OneshotVC} and vector quantization~\cite{wu2020one, wu2020vqvc+}, that are effective in disentangling speaker from the content.

%The  factor  of  speaker  plays  an important role  in  voice conversion.  For  example,  in  most  pair-wise voice conversion frameworks such as CycleGAN \cite{cycle_gan}  (one source  and  one target),  speaker  identity  is  only responsible for designating a frame to the input (if it is from the source) or to the output (if otherwise). It is useful to build a voice  conversion framework without explicitly  exploiting speaker-dependent factors. 
\begin{figure}[t]
  \centering
  \includegraphics[scale=0.62]{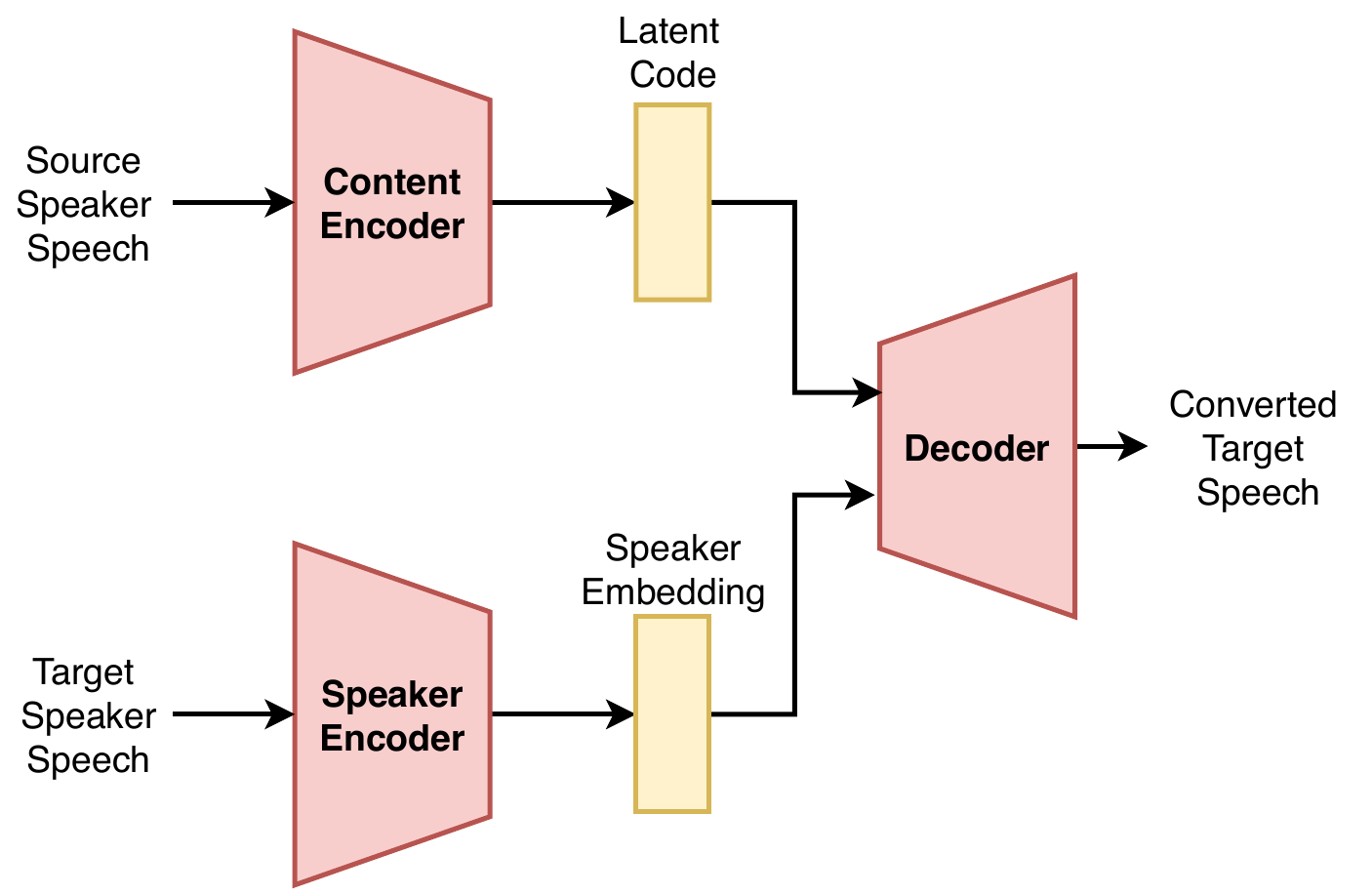}
 \caption{A typical auto-encoding network for voice conversion, where the encoders and decoder learn to disentangle speaker from linguistic content. At run-time, the linguistic content of the source speech represented by latent code and speaker embedding of a target speaker are combined to generate target speech.}
 \label{section5c-d}
\end{figure}
An auto-encoder learns to reproduce its input as its output. Therefore, parallel training data is not required. An encoder learns to represent the input with a latent code, and a decoder learns to reconstruct the original input from the latent code. The latent code can be seen as an information bottleneck which, on one hand, lets pass information necessary, e.g. speaker independent linguistic content, for  perfect  reconstruction, and  on the other hand, forces some information to be discarded, e.g. speaker, noise and channel information~\cite{hsu1}. Variational  auto-encoder (VAE) \cite{kingma2013auto} is the stochastic version of auto-encoder, in which the encoder produces distributions over latent representations, rather than deterministic latent codes, while the decoder is trained on samples from these  distributions. Variational auto-encoder is more suitable than deterministic auto-encoder in synthesizing new samples.

 Chorowski et al. \cite{wavenet-vae2019} provide a comparison of three auto-encoding neural networks by studying how they learn a representation from speech data to separate speaker identity from the linguistic content. It was shown that discrete representation, that is the latent code obtained from VQ-VAE \cite{van2017neural, razavi2019generating}, preserves  the  most  linguistic content while  also  being  the  most  speaker-invariant. Recently, a group latent embedding technique for VQ-VAE is studied to improve the encoding process, which divides the embedding dictionary into groups and uses the weighted average of atoms in the nearest group as the latent embedding \cite{ding2019group}.  
 
 The concept of a VAE-based voice conversion framework ~\cite{Hsu2016} can be illustrated in Figure 10. 
The  decoder  reconstructs  the  utterance  by  conditioning on the latent code extracted by the encoder, and separately on a speaker code, which could be an one-hot vector~\cite{Hsu2016,huang2018voice} for a close set of speakers, or an i-vector\cite{dehak2010front}, bottleneck speaker representation\cite{li2018many}, or d-vector\cite{saito2018non} for an open set of speakers. By explicitly conditioning the decoder on speaker identity,  the encoder is forced to capture speaker-independent information in the latent code from a multi-speaker database.

%It   removes   the   requirement   of   parallel   corpora   or   phonetic alignments  to  train  a  spectral  conversion  system. We note that this proposed idea uses a one-hot  vector for speaker representation, hence the framework only works for particular source and target speakers. 

%{(This paragraph needs to be merged with others. -Haizhou ) In a many-to-many voice conversion~\cite{huang2018voice}, it was studied to replace the one-hot vector \cite{Hsu2016, saito2018non} with speaker embedding, derived from a speaker encoder, such as d-vectors, and i-vectors \cite{saito2018non}, \cite{kinnunen2017non}. Once the  speaker representation  of  the  unknown  target  speaker is  obtained  from  the  limited  speech \cite{li2018many},  it  is  likely  that  the decoder  can  blend  speaker  and  phonetic  representations  to synthesize a speaker-dependent spectral frame, thus achieving M2M conversion. }

Just like other auto-encoder, VAE decoder tends to generate over-smoothed speech. This can be problematic for voice conversion because
the network may generate poor quality buzzy-sounding speech. Generative adversarial networks (GANs)~\cite{Huang_2020VAE-GAN} were proposed as one of the solutions to the over-smoothing problem \cite{kaneko2017sequence}. GANs offer a general framework for training a data generator
in such a way that it can deceive a real/fake discriminator that attempts to distinguish real data and fake data produced by the generator \cite{zhang2019self, liu2016coupled, zou2020edge}. By incorporating the GAN concept into VAE, VAE-GAN was studied for voice conversion with non-parallel training data~\cite{hsu3} and in cross-lingual voice conversion~\cite{sismanstudy}.
%With this method, the adversarial loss designed using a GAN discriminator is incorporated into the training loss to make the decoder outputs of a CVAE \cite{saito2018non} indistinguishable from real speech features. 
It was shown that VAE-GAN~\cite{Huang_2020VAE-GAN} produces more natural sounding speech than the standard VAE method~\cite{Hsu2016, li2018many}.

 A recent study on sequence-to-sequence non-parallel voice conversion~\cite{liu2020transferring} shows that it is possible to explicitly model the transfer of other aspects of speech, such as source rhythm, speaking style, and emotion to the target speech. 

\section{Evaluation of Voice Conversion}

Effective quality assessment of voice quality is required to validate the algorithms, to measure the technological progress, and to benchmark a system against the state-of-the-art. Typically, we report the results in terms of objective and subjective measurements. 

To provide an objective evaluation, a reference speech is required. The common objective evaluation metrics include Mel-cepstral distortion (MCD) \cite{mcd1} for spectrum, and PCC \cite{benesty2009pearson} and RMSE \cite{chai2014root, willmott2005advantages, grancharov2008speech} for prosody.  We note that, such metrics are not always correlated with human perception partly because they measure the distortion of acoustic features rather than the waveform that humans actually listen to.

Subjective evaluation metrics, such as the mean opinion score (MOS) \cite{streijl2016mean, chu2008optimization, viswanathan2005measuring, kain1998spectral}, preference tests \cite{kain2001design, Wu2014} and best-worst scaling \cite{Flynn2014} could represent the intrinsic naturalness and similarity to the target. We note that, for subjective evaluation to be meaningful, a large number of listeners are required, that is not always possible in practice.

%An interesting research direction is to see if we are able to evaluate converted speech with deep neural networks. For example, MOSNet \cite{lo2019mosnet} is a deep learning-based assessment tool for voice conversion that uses large-scale listening test results to train the model. Naturalness assessment by MOSNet achieves high correlation with human MOS ratings at the system level and fair correlation at the utterance level. Next we  briefly summarize objective and subjective evaluation techniques, as well as deep learning for quality assessment. 

\subsection{Objective Evaluation}
\subsubsection{Spectrum Conversion}
To provide an objective evaluation, first of all, we need a reference utterance spoken by the target speaker. Ideally the converted speech is very close to the reference speech. We can measure the differences between them by comparing their spectral distances. However, there is no guarantee that the converted speech and the reference speech is of the same length. In this case, a frame aligner is required to establish the frame-level mapping. 

Mel-cepstral distortion (MCD) \cite{mcd1} is commonly used to measure the difference between two spectral features~\cite{berrak-journal, toda2016voice, zhang2020deepconversion, lai2016phone}. It is calculated  between the converted and target Mel-cepstral coefficients, or MCEPs, \cite{black2012articulatory, logan2000mel}, $\hat{\textbf{y}}$ and $\textbf{y}$. Suppose that each MCEP vector consists of 24 coefficients, we have $\hat{\textbf{y}} = \{ m_{k,i}^{c} \}$ and ${\textbf{y}} = \{ m_{k,i}^{t} \}$ at frame $k$, where  $i$ denotes the $i$th coefficient in the converted and target MCEPs. 
\begin{equation}
    MCD[dB]=\frac{10}{\ln{10}}\sqrt{2\sum_{i=1}^{24}(m_{k,i}^{t}-m_{k,i}^{c})^2}
\end{equation}

We note that a lower MCD indicates better performance. However, MCD value is not always correlated with human perception. Therefore, subjective evaluations, such as MOS and similarity score, are also conducted.  

\subsubsection{Prosody Conversion}
Speech prosody of an utterance is characterized by phonetic duration, energy contour, and pitch contour. To effectively measure how close the prosody patterns of converted speech is to the reference speech, we need to provide measurements for the three aspects. 

The alignment between the converted speech and the reference speech provides the information about how much the phonetic duration differs one another. We can derive the number of frames that deviate from the  ideal  diagonal  path on average, such as frame disturbance~\cite{chitra2017framedisturbance}, to report the differences of phonetic duration.

%% motivate prosody work here... 
%Thus far, voice conversion studies are mainly focused on the conversion of spectrum. However, speaker identity is also characterized by its prosody features, such as fundamental frequency (F0). %Computational modeling of prosody has been a challenging task for many reasons. For example, prosody is described at supra-segmental level while spectrum is at short-time frame; prosody consists of F0 and energy among others that can vary highly.  Prosody is also hierarchical in nature \cite{csicsman2017transformation, xu2011speech} and it can be affected by both short term as well as long term dependencies \cite{Sanchez2014}. Fundamental frequency (F0) is a crucial prosodic feature in speech, hence previous studies of prosody conversion mainly focus on transformation of F0 \cite{gillett2003transforming}.

Pearson Correlation Coefficient (PCC)~\cite{zhou2020transforming,berrak-journal} and Root Mean Squared Error (RMSE) have been widely used as the evaluation metrics to measure the linear dependence of prosody contours or energy contours between two speech utterances.

We next take the measurement of two prosody contours as an example. PCC between the aligned pair of converted and target F0 sequences is given as follows,
\begin{equation}
    \rho(F0^c,F0^t) = \frac{cov(F0^c,F0^t)}{\sigma_{F0^c}\sigma_{F0^t}} 
\end{equation}
where $\sigma_{F0^c}$ and $\sigma_{F0^t}$ are the standard deviations of the converted F0 sequences ($F0^c$) and the target F0 sequences ($F0^t$), respectively. We note that a higher PCC value represents better F0 conversion performance.  

The RMSE between the converted F0 and the corresponding target F0 is defined as,
\begin{equation}
    RMSE=\sqrt{\frac{1}{K}\sum_{k=1}^{K}(F0_{k}^c-F0_{k}^t)^2}
\end{equation}
where $F0_{k}^c$ and $F0_{k}^t$ denote the converted and target F0 features, respectively. $K$ is the length of $F0$ sequence, or the total number of frames. We note that a lower RMSE value represents better $F0$ conversion performance. The same measurement applies to energy contours as well. 

Other generally-accepted metrics for prosody transfer include F0 Frame Error (FFE) \cite{chu2009reducing} and Gross Pitch Error (GPE) \cite{nakatani2008method}. We note that GPE reports the percentage
of voiced frames whose pitch values are more than
20$\%$ different from the reference, while FFE reports the percentage of frames that either contain a 20$\%$ pitch error or a voicing decision error~\cite{skerry2018towards}. 

\subsection{Subjective Evaluation}
Mean Opinion Score (MOS) has been widely used in listening tests~\cite{berrak_is18, berrak-journal, berrak3,watanabe2002transformation, Kobayashi2016, Sun2015, ramani2014cross, turk2006robust, desai2010spectral}. In MOS experiments, listeners rate the quality of the converted voice using a 5-point scale: “5” for excellent, “4” for good, “3” for fair, “2” for poor, and “1” for bad. There are several evaluation methods that are similar to MOS, for example: 1) DMOS \cite{tamura1998speaker, grancharov2006low, wester2015we}, which is a “degradation” or “differential” MOS test, requiring listeners to rate the sample with respect to this reference, and 2) MUSHRA \cite{zielinski2007potential, benisty2011voice, vit2018analysis}, which stands for MUltiple Stimuli with Hidden Reference and Anchor, and requires fewer participants than MOS to
obtain statistically significant results. 

Another popular subjective evaluation is preference test, also denoted as AB/ABX test \cite{zhang2008text, kain1998spectral, Sun2015} or XAB test \cite{Toda2007, mizuno1995voice}. In AB tests, listeners are presented with two speech samples and asked to indicate which one has more of a certain property; for example in terms of naturalness, or similarity. In ABX test, similar to that of AB, two samples are given but an extra reference sample is also given. Listeners need to judge if A or B more like X in terms of naturalness, similarity, or even emotional quality \cite{zhou2020transforming}. In XAB test, listeners are presented the original target speech sample first, and then a pair of converted voices randomly. We note that it is not practical to use AB, ABX or XAB test for the comparison of many VC systems at the same time. MUSHRA is another type of voice quality test in telecommunication~\cite{recommendation20011534}, where the reference natural speech and several other converted samples of the same content are presented to the listeners in a random order. The listeners are asked to rate the speech quality of each sample between 0 and 100.  

It is known that people are good at picking the extremes but their preferences for anything in between might be fuzzy and inaccurate when presented with a long list of options. Best-Worst Scaling (BWS)~\cite{Flynn2014} is proposed for voice conversion quality assessment~\cite{ccicsman2017sparse}, where listeners are presented only with a few randomly selected options each time. With many such BWS decisions, Best-Worst Scaling can handle a long list of options and generates more discriminating results, such voice quality ranking, than MOS and preference tests.  

We note that subjective measures can represent the intrinsic naturalness and similarity of a voice conversion system. However, such evaluation can be time-consuming and expensive as they involve a large number of listeners.

\subsection{Evaluation with Deep Learning Approaches}
%The evaluation of voice conversion reports both objective and subjective measurements. Objective measures such as the Mel-cepstral distance (MCD) and subjective measures such as mean opinion score (MOS) and similarity score are widely used for automatically measuring the quality of converted speech. 
%While objective and subjective evaluation metrics are good indicators of system performance, the objective metrics don't always correlate with human perception. On the other hand, human listening tests can be time-consuming and expensive as they involve a large number of participants. Automatic assessment of voice quality has become a research topic by itself.  
The study of perceptual quality evaluation seeks to approximate human judgement with computational models of psychoacoustic motivation. It provides insights into how humans perceive speech quality in listening tests, and suggests assessment metrics, that are required in speech communication, speech enhancement, speech synthesis,  voice conversion and any other speech production or transmission applications.  Perceptual Evaluation of Speech Quality (PESQ)~\cite{rix2001perceptual} is an ITU-T recommendation that is widely used as industry standard. It provides objective speech quality evaluation that predicts the human-perceived speech quality. 

However, the PESQ formulation requires the presence of reference speech, that considerably restricts its use in voice conversion applications, and motivates the study of perceptual evaluations without the need of reference speech. Those metrics that don't require reference speech are called non-intrusive evaluation metrics. For example, Fu et al.~\cite{fu2018quality} propose Quality-Net \cite{fu2018quality} that is an end-to-end model to predict PESQ ratings, that are the proxy for human ratings. Yoshimura et al.~\cite{yoshimura2016hierarchical}, Patton et al.~ \cite{patton2016automos} propose a CNN-based naturalness predictor to predict human MOS ratings, among other non-intrusive assessment metrics \cite{cernak2005evaluation, huang2011prediction, remes2013objective}.

%We note that with these methods, it is difficult to get a high correlation with human utterance-level ratings as the listening test is subjective and listeners may provide variant ratings to the same utterance. These methods have shown the capability of neural networks in modeling human perception for enhanced synthetic speech, and open a research direction for voice conversion community. 

Lo et al.~\cite{lo2019mosnet} propose MOSNet, another non-intrusive assessment technique based on deep neural networks, that learns to predict human MOS ratings. MOSNet scores are highly correlated with  human  MOS  ratings  at  system level, and fairly correlated at utterance level. While it is a non-intrusive evaluation metric for naturalness, MOSNet can also be modified and re-purposed to predict the similarity scores between target speech and converted speech. It provides similarity scores with fair correlation values to human ratings on VCC 2018 dataset. MOSNet marks a recent  advancement towards automatic perceptual quality evaluation~\cite{williams2020comparison}, which is free and open-source.

Last but not least, both Frechet DeepSpeech Distance (FDSD, cFDSD) and Kernel DeepSpeech Distance (KDSD, cKDSD) have been found to be well correlated with MOS for speech generation \cite{binkowski2019high}. We note that Frechet DeepSpeech Distance is motivated by Frechet Inception Distance (FID) \cite{heusel2017gans}, whereas Kernel DeepSpeech Distance is motivated by Kernel Inception Distance (KID) \cite{binkowski2018demystifying}. In both of these frameworks, the Inception image recognition network has been replaced with the DeepSpeech audio recognition network for evaluation of speech generation.

%and shows high . It is basically a mean opinion score (MOS) predictor that is trained with MOS evaluations of VCC 2018 as the ground truth, and tested on VCC 2016 data. Therefore, it is seen as an end-to-end objective assessment model for voice conversion. Experiments show that the predicted scores of the proposed MOSNet are highly correlated with human MOS ratings at the system level while being fairly correlated with human MOS ratings at the utterance level. MOSNet can be beneficial for voice conversion community as it reduces the need for expensive human rating \cite{williams2020comparison}. 

%We note that the best performance of MOSNet reported for similarity prediction achieves an accuracy of 69.6 \%, while it achieves sufficiently high correlation of 0.917 for MOS prediction. MOSNet considers MOS prediction as a regression task, and uses the traditional MSE-based objective functions. We believe that a further study on human perception theory can effectively improve the training of MOSNet for similarity prediction.

\section{Voice Conversion Challenges}

In this section, we would like to give an overview of the series of voice conversion challenges, that provide shared tasks with  common data sets and evaluation metrics for fair comparison of algorithms.  The voice conversion challenge (VCC) is a biannual event since 2016. In a challenge, a common database is provided by the organizers.  The participants build voice conversion systems using their own technology, and the organizers evaluate the performance of the converted speech. The main evaluation methodology is a listening test in which crowd-sourced evaluators rank the naturalness and speaker similarity.

The 2016 challenge offers a standard voice conversion task using a parallel training database was adopted \cite{toda2016voice}. The 2018 challenge features a more advanced conversion scenario using a non-parallel database~\cite{Lorenzo-Trueba2018}. The 2020 challenge puts forward a cross-lingual voice conversion research problem. A summary of VCC 2016, VCC 2018 and VCC 2020 is also provided in Table I.

\subsection{Why is the Challenge Needed?}

As described earlier, many of the voice conversion approaches are data-driven, hence speech data are required to train models and for conversion evaluation. To compare such data-driven methods each other precisely, a common database that specifies training and evaluation data explicitly is needed. However, such  common database did not exist until 2016. % and hence various speech databases were used in inconsistent ways. 
Without common databases, researchers have to re-implement others' system with their own databases before trying any new ideas. In such situation, it is not guaranteed that the re-implemented  system achieves the expected performance in the original work. 

To address the same problem, the TTS community gave birth to the first Blizzard challenge in 2005. Since then, the challenge has defined various standard databases for TTS and has made comparisons of TTS much fairer and easier. The motivations of VCC are exactly the same as those of the Blizzard challenges. VCC introduced a few standard databases for voice conversion and also defined the common training and evaluation protocols.  All the converted speech submitted by the participants for the challenges have been released publicly. In this way,  researchers can compare the performance of their voice conversion system with that of other state-of-the-art systems without the need of re-implementation.

Another need on voice conversion standard databases arose from biometric speaker recognition community. As the voice conversion technology could be misused for attacking speaker verification systems, anti-spoofing countermeasures are required~\cite{WU2015130}. This is also called presentation attack detection.  Anti-spoofing techniques aim at discriminating between fake artificial inputs presented to biometric authentication systems and genuine inputs.  If sufficient knowledge and data regarding the spoofed data is available, a binary classifier can be constructed to reject artificial inputs. %If such knowledge and data are not available, one-class classifier techniques such as anomaly detection and outlier detection are used instead. 
Therefore, the common VCC databases are also important for anti-spoofing research.  With many converted speech data from advanced voice conversion systems, researchers in the biometric community can develop anti-spoofing models to strengthen the defence of speaker recognition systems, and to evaluate their vulnerabilities. 
\begin{table*}
\centering
\begin{tabular}{cccccc}
 \hline
 \textbf{Challenge}& \textbf{Language} &\textbf{Task} & \textbf{Training Data}&\textbf{ \# Speakers }&  \textbf{Testing Data} \\ [0.5ex] 
\hline \newline
\multirow{1}{*}{VCC 2016} 
& monolingual & parallel  & 162 paired utterances & 4 source, 4 target & 54 utterances \\  
 \hline \hline
 
\multirow{2}{*}{VCC 2018} 
& monolingual & parallel  & 81 paired utterances & 4 source, 4 target & 35 utterances \\  
& monolingual & nonparallel  & 81 unpaired utterances & 4 source, 4 target & 35 utterances \\  

 \hline \hline
\multirow{2}{*}{VCC 2020} 
& monolingual& parallel + nonparallel & 20 paired, 50 unpaired utterances & 4 source, 4 target & 25 utterances \\ 
& crosslingual & nonparallel & 70 unpaired utterances & 4 source, 6 target & 25 utterances  \\ 
\hline

\end{tabular}

\caption {Summary of VCC 2016, VCC 2018 and VCC 2020.}
\end{table*}

\subsection{Overview of the 2016 Voice Conversion Challenge}

We first overview the 2016 voice conversion challenge \cite{toda2016voice} and its datasets\footnote{The VCC2016 dataset is available at \url{https://doi.org/10.7488/ds/1575}}. As the first shared task in voice conversion, a parallel voice conversion task and its evaluation protocol are defined for VCC 2016. The parallel dataset consists of 162 common sentences uttered by both source and target speakers. Target and source speakers are four native speakers of American English (two females and two males), respectively. In the challenge, the participants develop the conversion systems and  produce converted speech for all possible source-target pair combinations. In total, eight speakers (plus two unused speakers) are included in the VCC 2016 database. The number of test sentences for evaluation is 54. 

The main evaluation methodology adopted for the ranking is subjective evaluation on perceived naturalness and speaker similarity of the converted samples to target speakers. The naturalness is evaluated using the standard five-point scale mean-opinion score (MOS) test ranging from 1 (completely unnatural) to 5 (completely natural). The speaker similarity was evaluated using the Same/Different paradigm \cite{Wester+2016}. Subjects are asked to listen to two audio samples and to judge if they are speech signals produced by the same speaker in a four point scale: ``Same, absolutely sure'', ``Same, not sure'', ``Different, not sure'' and ``Different, absolutely sure.'' As the perceived speaker similarity to a target speaker, and the perceived voice quality are not necessarily correlated, it is important to use a scatter-plot to observe the trade-off between the two aspects.

In the 2016 challenge, 17 participants submitted their conversion results. Two hundreds native listeners of English joined the listening tests. It is reported that the best system using GMM and waveform filtering obtained an average of 3.0 in the five-point scale evaluation for the naturalness judgement, and about 70\% of its converted speech samples are judged to be the same as target speakers by listeners. However, it is also confirmed that there is still a huge gap between target natural speech and the converted speech. We observe that it remains a unsolved challenge to achieve good quality and speaker similarity at that time. More details of VCC 2016 can be found at \cite{Wester+2016}.  Details of best performing systems are reported in \cite{Kobayashi2016}.

\subsection{Overview of the 2018 Voice Conversion Challenge}

Next we give an overview of the 2018 voice conversion challenge~\cite{Lorenzo-Trueba2018} and its datasets\footnote{The VCC2018 dataset is available at \url{https://doi.org/10.7488/ds/2337}.}. VCC 2018 offers two tasks, parallel and non-parallel voice conversion tasks. A dataset and its evaluation protocol are defined for each task. The dataset for the parallel conversion task is similar to that of the 2016 challenge, except that it has a smaller number of common utterances uttered by source and target speakers. Target and source speakers are four native speakers of American English (two females and two males), respectively, but, they are different speakers from those used for the 2016 challenge. Like the 2016 challenge, the participants were asked to develop conversion systems and to produce converted data for all possible source-target pair combinations. 

VCC 2018 introduced a non-parallel voice conversion task for the first time. %The dataset for the non-parallel task has exactly the same target speaker's data. 
The same target speakers' data in the parallel task are used as the target. However, the source speakers are four native speakers of American English (2 females and 2 males) different from those of the parallel conversion task and their utterances are also all different from those of the target speakers. Like the parallel voice conversion task, converted data for all possible source-target pair combinations needed to be produced by the participants. In total twelve speakers are included in the VCC 2018 database. Each of the source and target speakers has a set of 81 sentences as training data, which is half of that for VCC 2016. The number of test sentences for evaluation is 35. 

%The main evaluation methodology adopted for the ranking is subjective evaluation on perceived naturalness and speaker similarity of the converted samples to target speakers. The naturalness was evaluated using the standard five-point scale mean-opinion score (MOS) test ranging from 1 (completely unnatural) to 5 (completely natural). The speaker similarity was evaluated using the Same/Different paradigm \cite{Wester+2016}. Subjects were asked to listen to two audio samples and to judge if they are speech signals produced by the same speaker in a four point scale: ``Same, absolutely sure'', ``Same, not sure'', ``Different, not sure'' and ``Different, absolutely sure.'' Appropriate conversion process improves perceives speaker similarity to a target speaker, but it typically degrades perceived quality due to conversion artifacts. Therefore it is important to use a scatter-plot matching naturalness and similarity scores and see trade-off between the two aspects, too. 

In the 2018 challenge, 23 participants submitted their conversion results to the parallel conversion task, with 11 of them additionally participating in the non-parallel conversion task. The same evaluation methodology as the 2016 challenge was adopted for the 2018 challenge and 260 crowd-sourced native listeners of English have joined the listening tests. It was reported that in both tasks, the best system using phone encoder and neural vocoder obtained an average of 4.1 in the five-point scale evaluation for the naturalness judgement and about 80\% of its converted speech samples were judged to be the same as target speakers by listeners. It was also reported that the best system has similar performance in both the parallel and non-parallel tasks in contrast to results reported in literature. 

In VCC 2018, the spoofing countermeasure was introduced as an supplement to subjective evaluation of voice quality, that brought together the voice conversion and speaker verification research community. More details of the 2018 challenge can be found at \cite{Lorenzo-Trueba2018}. Details of best performing systems are reported in \cite{Wu2018,Liu2018}. 

From this challenge, we observed that new speech waveform generation paradigms such as WaveNet and phone encoding have brought significant progress to the voice conversion field.  Further improvements have been achieved in the follow up papers \cite{8607053,8936924} and new VC systems that exceed the challenge's best performance have already been reported.

\subsection{ Overview of the 2020 Voice Conversion Challenge}
%\subsection{Overview of the 2020 Voice Conversion Challenge} 

The 2020 voice conversion challenge \cite{zhao2020voice} consists of two tasks: 1) non-parallel training in the same language (English); and 2) non-parallel training over different languages (English-Finnish, English-German, and English-Mandarin). 

In the first task, each participant trains voice conversion models for all source and target speaker pairs using up to 70 utterances, including 20 parallel utterances and 50 non-parallel utterances in English, for each speaker as the training data. Overall, 16 voice conversion models (i.e., 4 sources by 4 targets) are to be developed. In the second task, each participant develops voice conversion models for all source and target speaker pairs using up to 70 utterances for each speaker (i.e., in English for the source speakers, and in Finnish, German, or Mandarin for the target speakers) as the training data. Overall, 24 conversion systems (i.e., 4 sources by 6 targets) are to be developed. 

In the 2020 challenge, 31 participants submitted their results to the first task, and 28 participants submitted their results to the second task.  The participants were allowed  to mix and combine different source speaker's data to train speaker-independent models. Moreover, the participants can also use orthographic transcriptions of the released training data to develop their voice conversion systems. Last but not least, the participants were free to perform manual annotations of the released training data, which can effectively improves the quality of the converted speech.  

The 2020 challenge organizers also built several baseline systems including the top system of the previous challenge on the new database. The codes of CycleVAE-based baseline\footnote{ \url{https://github.com/bigpon/vcc20_baseline_cyclevae}} and  Cascade ASR + TTS based VC \footnote{\url{ https://github.com/espnet/espnet/tree/master/egs/vcc20}.} are released so that participants can build the basic systems easily and focus on their own innovation.  The 2020 challenge  features a multifaceted evaluation. In addition to the traditional evaluation metrics, the challenge also reports the speech recognition, speaker recognition, and anti-spoofing evaluation results on the converted speech. 

According to the final report, it was encouraging to see that the  speaker  similarity  scores  of  several systems are very close to that of natural speech of target speakers in the first task. However, none of the systems achieved human-level naturalness. The second task is a  more  challenging  one. While we observed lower  overall  naturalness  and similarity scores than those of the first task, the MOS scores of the best systems were higher than 4.0.

%still ongoing for the time being, and the results will be released on July 31st, 2020. 

\subsection{Relevant Challenges -- ASVspoof Challenge}

The spoofing capability against automatic speaker verification is a related topic to voice conversion, that has also been organized as technology challenges. 
The ASVspoof series of challenges are such biannual events, which started in 2013. Like in the voice conversion challenges, the organizers release a common database including many pairs of spoofed audio (converted, generated audio or replay audio) and genuine audio to the participants, who build anti-spoofing models using their own technology. The organizers rank the detection accuracy of the anti-spoofing results submitted by the participants. 

In 2015, the first anti-spoofing database including various types of spoofed audio using voice conversion and TTS systems was constructed. This database became a reference standard in the automatic speaker verification (ASV) community \cite{Wu-ASVspoof2015,7858696}. The main focus of the 2017 challenge was a replay task, where a large quantity of real-world replay speech data were collected \cite{Kinnunen2017-assessing}. In 2019, an even larger database including converted, generated, and replay speech data was constructed \cite{Todisco2019}. The best performing systems in the 2016 and 2018 voice conversion challenges were also used for generating advanced spoofed audio \cite{wang2019asvspoof}. The challenges revealed that some anti-spoofing systems outperform human listeners in detecting spoofed audio. 

\section{Resources}

In addition to the voice conversion challenge databases described above, the CMU-Arctic database \cite{kominek2004cmu} and the VCTK databases \cite{veaux2016vctk} are also popular for voice conversion research.  The current version of the CMU-Arctic database\footnote{\url{http://www.festvox.org/cmu_arctic/}} has 18 English speakers and each of them reads out the same set of around 1,150 utterances, which are carefully selected from out-of-copyright texts from Project Gutenberg. This is suitable for parallel voice conversion since sentences are common to all the speakers. The current version (ver.\ 0.92) of the CSTR VCTK corpus\footnote{\url{https://doi.org/10.7488/ds/2645}} has speech data uttered by 110 English speakers with various dialects. Each speaker reads out about 400 sentences, which are selected from  newspapers, the rainbow passage and an elicitation paragraph used for the speech accent archive. Since the rainbow passage and an elicitation paragraph are common to all the speakers, this database can be used for both parallel and non-parallel voice conversion. 

Since neural networks are data hungry and generalization to unseen speakers is a key for successful conversion, large-scale, but, low-quality databases such as LibriTTS and VoxCeleb are also used for training some components required (e.g.\ speaker encoder) for voice conversion. The LibriTTS corpus \cite{Zen2019} has 585 hours of transcribed speech data uttered by total of 2,456 speakers. The recording condition and audio quality are less than ideal, but, this corpus is suitable for training speaker encoder networks or generalizing any-to-any speaker mapping network. The VoxCeleb database \cite{NAGRANI2020101027} is further a larger scale speech database consisting of about 2,800 hours of untranscribed speech from over 6,000 speakers. This is an appropriate database for training noise-robust speaker encoder networks. 

There are many open-source codes for training VC models. For instance, spocket \cite{Kobayashi2018} supports GMM-based conversions and ESPnet \cite{Watanabe2018} supports cascaded ASR and TTS system. In addition, there are many open-source codes for neural-network based voice conversion written by the community at github\footnote{\url{https://paperswithcode.com/task/voice-conversion}}.

%Neural vocoders??  

\section{Conclusion}
This article provides a comprehensive overview of the voice conversion technology, covering the fundamentals and practice till August 2020. We reveal the underlying technologies and their relationship from the statistical approaches to deep learning, and discuss their promise and limitations. We also study the evaluation techniques for voice conversion. Moreover, we report the series of voice conversion challenges and resources that are useful information for researchers and engineers to start voice conversion research.   

\section{Acknowledgment}
The work by Berrak Sisman is supported by SUTD Start-up Grant Artificial Intelligence for Human Voice  Conversion  (SRG ISTD  2020  158)  and  SUTD  AI  Grant  titled `The Understanding and Synthesis of Expressive Speech by AI' (PIE-SGP-AI-2020-02).

The work by Haizhou Li is supported by the National Research Foundation, Singapore under its AI Singapore Programme (AISG Award No: AISG-GC-2019-002 and AISG Award No:  AISG-100E-2018-006). This research is also supported by Human-Robot Interaction Phase 1 (Grant No.: 192 25 00054), National Research Foundation, Singapore under the National Robotics Programme, and Programmatic Grant No.: A18A2b0046 (Human Robot Collaborative AI for AME) and A1687b0033 (Neuromorphic Computing) from the Singapore Government’s Research, Innovation and Enterprise 2020 plan in the Advanced Manufacturing and Engineering domain.

\bibliographystyle{IEEEbib}
\bibliography{berrak2}

\begin{IEEEbiography}[{\includegraphics[width=1in,height=1.25in,clip,keepaspectratio]{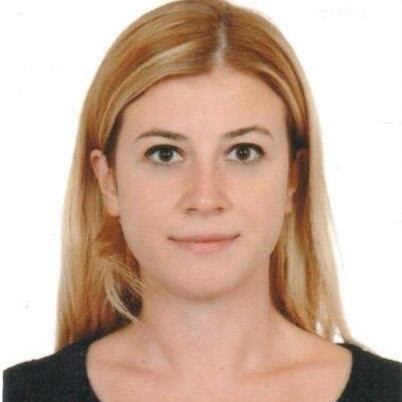}}]{Berrak Sisman} received the Ph.D. degree from National University of Singapore in 2020, fully funded by A*STAR Graduate Academy. She is currently an Assistant Professor at Singapore University of Technology and Design (SUTD). She is also an Affiliated Researcher and Team Leader at the National University of Singapore (NUS). She was an exchange PhD student at the University of Edinburgh and a visiting scholar at The Centre for Speech Technology Research, University of Edinburgh in 2019. She was attached to RIKEN Advanced Intelligence Project, Japan in 2018.  Her research interests include speech information processing, machine learning, speech synthesis and voice conversion. She has published in leading journals and conferences, including IEEE/ACM Transactions on Audio, Speech and Language Processing, ASRU, INTERSPEECH and ICASSP. %She has served as the Local Arrangement Co-chair of IEEE ASRU 2019, Chair of Young Female Researchers Mentoring @ASRU2019, and Chair of the INTERSPEECH Student Events in 2018 and 2019. 
\end{IEEEbiography}
  
\begin{IEEEbiography}[{\includegraphics[width=1in,height=1.25in,clip,keepaspectratio]{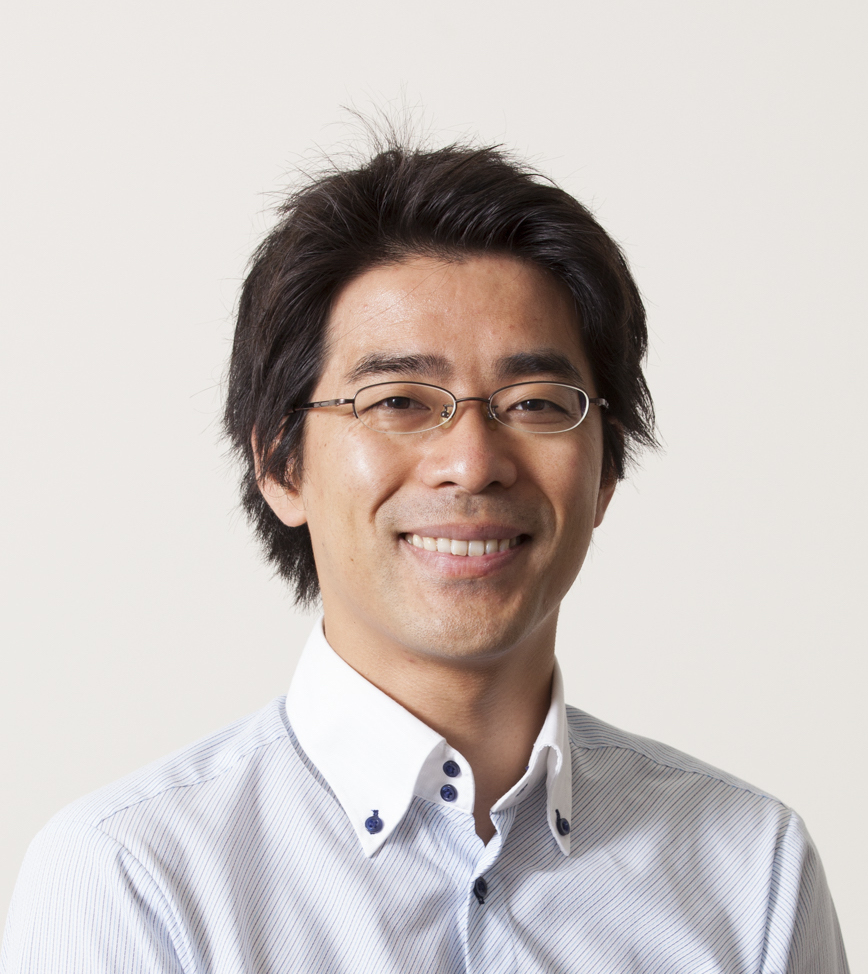}}]{Junichi Yamagishi} received the Ph.D.\ degree from the Tokyo Institute of Technology (Tokyo Tech), Tokyo, Japan, in 2006. %His Ph.D.\ dissertation pioneered speaker-adaptive speech synthesis. 
He is currently a Professor with the National Institute of Informatics, Tokyo, Japan, and also a Senior Research Fellow with The Centre for Speech Technology Research, The University of Edinburgh, Edinburgh, UK. Since 2006, he has authored or co-authored over 250 refereed papers in international journals and conferences. 
%He is a member of the Speech and Language Technical Committee. 
Prof.\ Yamagishi was a recipient of the Tejima Prize as the best Ph.D.\ thesis of Tokyo Tech in 2007. He received the Itakura Prize from the Acoustic Society of Japan in 2010, the Kiyasu Special Industrial Achievement Award from the Information Processing Society of Japan in 2013, the Young Scientists' Prize from the Minister of Education, Science and Technology in 2014, the JSPS Prize from the Japan Society for the Promotion of Science in 2016, and the 17th DOCOMO Mobile Science Award from the Mobile Communication Fund, Japan in 2018. He was one of the organizers for special sessions on Spoofing and Countermeasures for the  Automatic Speaker Verification at INTERSPEECH 2013, the 1st/2nd/3rd ASVspoof Evaluation, the Voice Conversion Challenge 2016/2018/2020, and the VoicePrivacy Challenge 2020. He was an Associate Editor of the IEEE/ACM Transactions on Audio, Speech, and Language Processing, a Lead Guest Editor of the IEEE Journal of Selected Topics in Signal Processing Special Issue on Spoofing and Countermeasures for Automatic Speaker Verification, and a member of the Technical Committee of the IEEE Signal Processing Society Speech and Language. He is now the Chairperson of ISCA Special Interest Group: Speech Synthesis (SynSig), a member of the Technical Committee for the Asia-Pacific Signal and Information Processing Association Multimedia Security and Forensics, an IEEE Senior Area Editor of the IEEE/ACM Transaction on Audio, Speech, and Language Processing.

\end{IEEEbiography}
\begin{IEEEbiography}[{\includegraphics[width=1in,height=1.25in,clip,keepaspectratio]{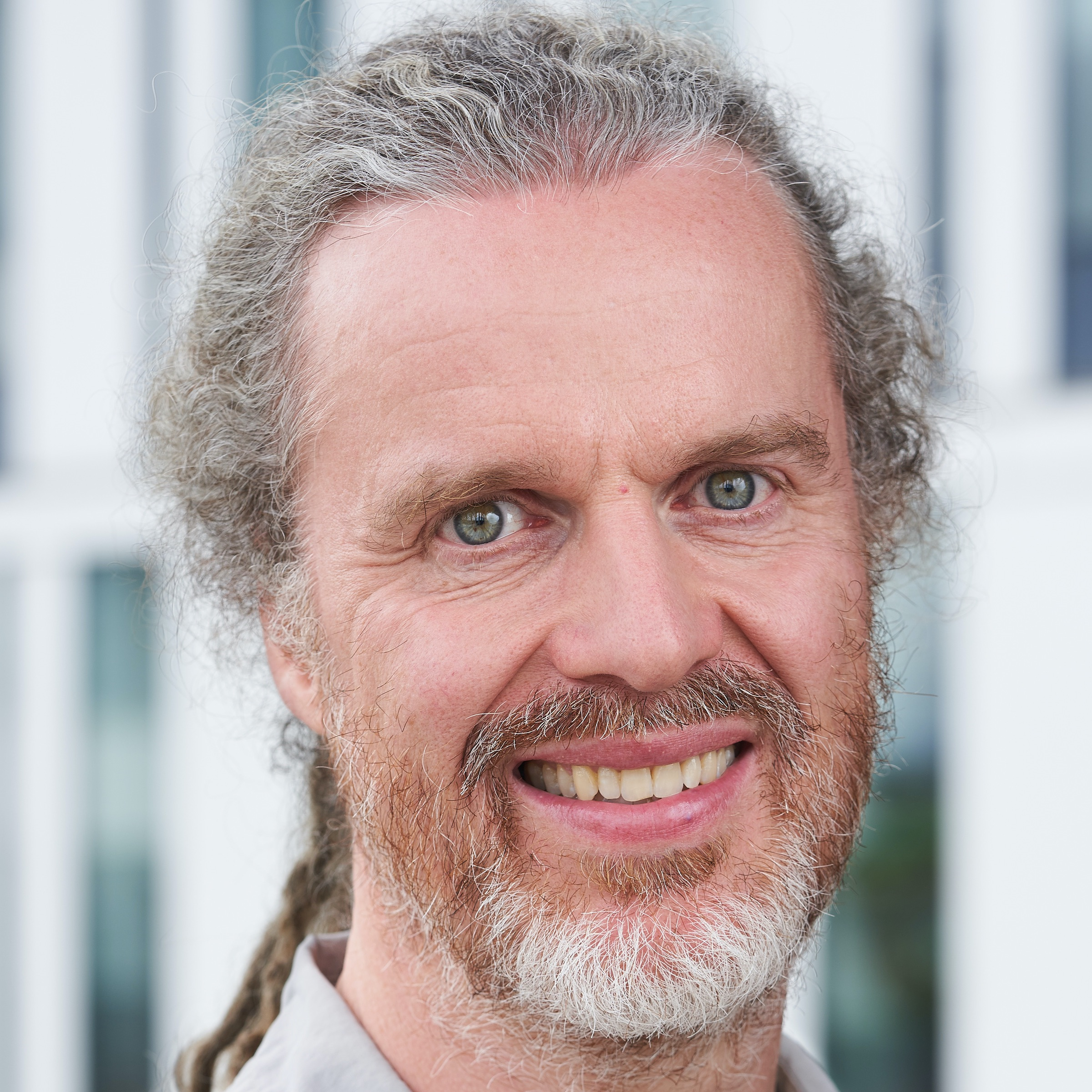}}]{Simon King} (M’95–SM’08–F’15) received the M.A. (Cantab) and M.Phil. degrees from the University of Cambridge, Cambridge, U.K., and the Ph.D. degree from University of Edinburgh, Edinburgh, U.K. He has been with the Centre for Speech Technology Research, University of Edinburgh, since 1993, where he is now Professor of Speech Processing and the Director of the Centre. His research interests include speech synthesis, recognition and signal processing and he has around 230 publications across these areas. He has served on the ISCA SynSIG Board and currently co-organises the Blizzard Challenge. He has previously served on the IEEE SLTC and as an Associate Editor of the IEEE/ACM Transactions on Audio, Speech, and Language Processing, and is currently an Associate Editor of Computer Speech and Language.

\end{IEEEbiography}
\begin{IEEEbiography}[{\includegraphics[width=1in,height=1.25in,clip,keepaspectratio]{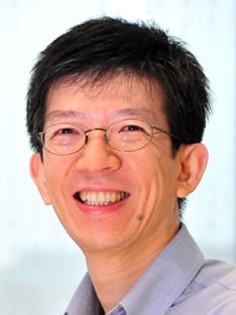}}]{Haizhou Li} (M’91-SM’01-F’14) received the B.Sc., M.Sc., and Ph.D degree in electrical and electronic engineering from South China University of Technology, Guangzhou, China in 1984, 1987, and 1990 respectively. Dr Li is currently a Professor at the Department of Electrical and Computer Engineering, National University of Singapore (NUS). His research interests include automatic speech recognition, speaker and language recognition, and natural language processing. Prior to joining NUS, he taught in the University of Hong Kong (1988-1990) and South China University of Technology (1990-1994). He was a Visiting Professor at CRIN in France (1994-1995), Research Manager at the Apple-ISS Research Centre (1996-1998), Research Director in Lernout \& Hauspie Asia Pacific (1999-2001), Vice President in InfoTalk Corp. Ltd. (2001-2003), and the Principal Scientist and Department Head of Human Language Technology in the Institute for Infocomm Research, Singapore (2003-2016). Dr Li served as  the Editor-in-Chief of IEEE/ACM Transactions on Audio, Speech and Language Processing (2015-2018), a Member of the Editorial Board of Computer Speech and Language (2012-2018), an elected Member of IEEE Speech and Language Processing Technical Committee (2013-2015), the President of the International Speech Communication Association (2015-2017), the President of Asia Pacific Signal and Information Processing Association (2015-2016), and the President of Asian Federation of Natural Language Processing (2017-2018). He was the General Chair of ACL 2012, INTERSPEECH 2014, and ASRU 2019. Dr Li is a Fellow of the IEEE and the ISCA. He was a recipient of the National Infocomm Award 2002 and the President’s Technology Award 2013 in Singapore. He was named one of the two Nokia Visiting Professors in 2009 by the Nokia Foundation, and U Bremen Excellence Chair Professor in 2019.

\end{IEEEbiography}
\end{document}